\documentclass[letterpaper,11pt]{article}
\usepackage{jheppub}
\usepackage{amsmath,amsthm,amssymb}
\usepackage{bm}
\usepackage{color}
\usepackage{graphicx}
\usepackage{caption,subcaption}
\usepackage{cancel}
\usepackage[dvipsnames]{xcolor}
\usepackage{mathrsfs}
\usepackage{amssymb}
\usepackage{dsfont}
\usepackage{float}
\usepackage{amsthm}
 \usepackage{physics}

\usepackage{float}
\usepackage{afterpage}
\usepackage{tikz}		% Basic tikz
\usetikzlibrary{cd}		% commutative diagrams
\usetikzlibrary{calc}		% Calculations in tikz
\usetikzlibrary{math} %needed tikz library for setting variables
\usetikzlibrary{intersections}
\usetikzlibrary{decorations.pathmorphing}
\usetikzlibrary{decorations.pathreplacing}
\usetikzlibrary{decorations.markings}

\newcommand*{\ZZ}{\mathbb{Z}}

\newcommand*{\RR}{\mathbb{R}}
\newcommand*{\CC}{\mathbb{C}}

\newcommand*{\target}{\mathcal{M}_{\mathrm{target}}}

\newcommand*{\hbu}{\mathcal{H}_\mathrm{BU}}
\newcommand*{\HH}{\mathrm{HH}}
\newcommand*{\BC}{\mathcal{BC}_1} % Collection of boundary conditions allowed for each connected component
\newcommand*{\ASU}{\mathcal{A}_{SU}} % Single-universe amplitude
\newcommand*{\tASU}{\widetilde{\mathcal{A}}_{SU}} % Single-universe amplitude
\newcommand{\piD}[1]{\mathcal{D}#1} % Path integral measure
\newcommand{\CPT}{{\sf  CPT} }

\definecolor{rust}{rgb}{0.8,0.2,0.2}
\usepackage{cleveref}
\crefname{section}{\S\!\!}{\S\S\!\!}
\Crefname{Section}{\S}{\S\S}
\crefname{appendix}{Appendix}{Appendices\!}
\crefname{figure}{Fig.\!}{Figs.\!}
\crefname{table}{Table \!}{Tables \!}

\newpage

\title{Baby Universes and Worldline Field Theories}

\author[a]{Eduardo Casali,}
\author[b]{Donald Marolf,}
\author[b]{Henry Maxfield,}
\author[a]{Mukund Rangamani}

\affiliation[a]{Center for Quantum Mathematics and Physics (QMAP)\\
Department of Physics \& Astronomy, University of California, Davis, CA 95616 USA
}

\affiliation[b]{Department of Physics, University of California, Santa Barbara, CA 93106, USA}

\emailAdd{ecasali@ucdavis.edu}
\emailAdd{marolf@physics.ucsb.edu}
\emailAdd{hmaxfield@physics.ucsb.edu}
\emailAdd{mukund@physics.ucdavis.edu}

\abstract{The quantum gravity path integral involves a sum over topologies that invites comparisons to worldsheet string theory and to Feynman diagrams of quantum field theory.  However,  the latter are naturally associated with the non-abelian algebra of quantum fields, while the former has been argued to define an abelian algebra of superselected observables associated with partition-function-like quantities at an asymptotic boundary.   We resolve this apparent tension by pointing out a variety of discrete choices that must be made in constructing a Hilbert space from such path integrals, and arguing that the natural choices for quantum gravity differ from those used to construct QFTs.  We focus on one-dimensional models of quantum gravity in order to make direct comparisons with worldline QFT.  We also restrict attention to models in which worldlines do not split apart or join together.
}

\begin{document}
\maketitle

%%%%%%%%%%%%%%%%%%%%%%%%%%%%%%%%%%%%%%%%%%%%%%%%%%%%%%%%%%%%

%~~~~~~~~~~~~~~~~~~~~~~~~~~~~~~~~~~~~~~~~~~~~~~~
\section{Introduction}
\label{sec:intro}
%~~~~~~~~~~~~~~~~~~~~~~~~~~~~~~~~~~~~~~~~~~~~~~

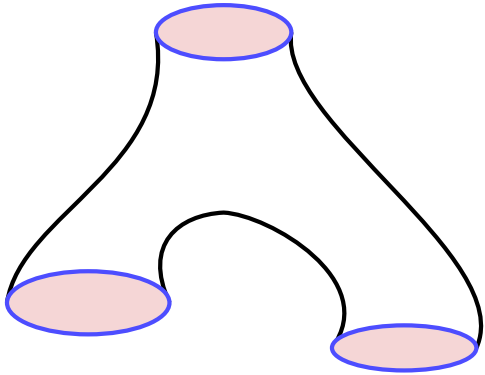
\begin{figure}[h]
\centering
\begin{tikzpicture}[scale=0.6]
\coordinate (u1) at (0,6);
\coordinate (u2) at (-3,0);
\coordinate (u3) at (4,-1);
\draw[ultra thick, black]  ($(u1) + (-1.5,0)$) .. controls (-1,3) and (-4.5,2) .. ($(u2) +(-1.8,0)$ );
\draw[ultra thick, black]  ($(u1) + (1.5,0)$) .. controls (1.2,4) and (6.5,1) .. ($(u3) +(1.6,0)$ );
\draw[ultra thick, black]   ($(u2) +(1.8,0)$ ) .. controls (-1.8,1.3) and (-1,1.95) .. ($(u1)+(0,-4)$) ..  controls (1, 1.95) and (3.5,0.5)  ..  ($(u3) +(-1.6,0)$ );
\draw[ultra thick, blue!70, fill=rust!20] (u1) ellipse [x radius = 1.5, y radius = 0.6];
\draw[ultra thick, blue!70, fill=rust!20] (u2) ellipse [x radius = 1.8, y radius = 0.7];
\draw[ultra thick,blue!70, fill=rust!20] (u3) ellipse [x radius = 1.6, y radius = 0.5];
\end{tikzpicture}
\caption{Schematic illustration of a spacetime wormhole connecting three boundaries.}
\label{fig:wormholes}
\end{figure}

Recent work has revived ideas  from the late 1980's and early 1990's \cite{Coleman:1988cy,Giddings:1988cx,Giddings:1988wv,Polchinski:1994zs} (based in turn on the earlier refs. \cite{Hawking:1987mz,Giddings:1987cg,Lavrelashvili:1987jg,Hawking:1988ae}) suggesting that spacetime wormholes elucidate the quantum physics of black holes
\cite{Saad:2018bqo,Saad:2019lba,Blommaert:2019wfy,Saad:2019pqd,Penington:2019kki,Marolf:2020xie,Blommaert:2020seb,Bousso:2020kmy,Stanford:2020wkf}, and other issues involving gravitational entropy \cite{Chen:2020tes}.   Here the term spacetime wormhole is used to denote a
configuration in the quantum gravity path integral for which two separate connected components of the spacetime boundary are connected through the bulk of spacetime; see  \cref{fig:wormholes}.

Furthermore, because boundary partition functions tend not to factorize in the presence of spacetime wormholes (see e.g., \cite{Maldacena:2004rf}), the above suggestions have led to renewed discussion of the nature of the AdS/CFT correspondence.  Indeed, as described in the above references it may be that a given bulk theory is dual to an ensemble of field theories -- though see also comments in \cite{Pollack:2020gfa}, \cite{McNamara:2020uza}, \cite{Belin:2020hea},  \cite{Liu:2020jsv} and \cite{Altland:2020ccq}.  The physics of spacetime wormholes and their possible implications are thus very much a part of current research.

However, attempting to understand issues involving spacetime wormholes brings one face-to-face with the absence of a fully developed and universally accepted set of rules for manipulating and interpreting quantum gravity path integrals.  This deficit can lead to much confusion in both the technical investigation of such  path integrals and in communicating the results.  Our work below seeks to aid both tasks by pointing out certain discrete choices that must be made in order to define a theory of quantum gravity from diagrams like those in \cref{fig:wormholes}, and by exploring certain implications of such choices.

In particular, one may note that the spacetimes of \cref{fig:wormholes} are similar to the diagrams of worldsheet string theory and to the Feynman diagrams of QFT.  Taken together with other parallels between free particles in Minkowski space and simple models of quantum gravity, this feature is often taken to suggest that the full structure of quantum gravity will again resemble that of string theory or QFT \cite{Kuchar1981,Caderni:1984pw,Moss:1986yd,McGuigan:1988vi,Banks:1988je,Rubakov:1988jf,Giddings:1988wv,Hosoya:1988aa,Strominger:1988si,Hawking:1991vs,Lyons:1991im}, in which context the process of building the corresponding theory of quantum gravity
is often called `third quantization' (also see  the recent discussion in \cite{Giddings:2020yes}).  However, we avoid using this term below due its status as a work-in-progress and the resulting lack of a clear definition in the literature.\footnote{For similar reasons, we also avoid use of the terms  which have previously  been applied, eg.,   ``multiverse field theory'' parenthetically mentioned in  \cite{Giddings:1988wv},  ``universal field theory'' of \cite{Banks:1988je}, and ``universe field theory"  coined in  \cite{Anous:2020lka}.}

We emphasize below that such QFT-like approaches correspond only to certain possible choices that might be made in interpreting the diagrams of  \cref{fig:wormholes}.  In contrast, distinctly different choices were (perhaps implicitly) made in recent comparisons of Jackiw-Teitelboim quantum gravity with ensembles of matrix models \cite{Saad:2018bqo,Saad:2019lba,Blommaert:2019wfy,Saad:2019pqd,Penington:2019kki,Blommaert:2020seb}, and (more explicitly) in general
arguments \cite{Marolf:2020xie} that spacetime wormholes lead to superselection sectors for boundary partition functions associated with states of a so-called `baby universe' Hilbert space;\footnote{Similar-but-different conclusions were reached in \cite{Coleman:1988cy,Giddings:1988cx,Giddings:1988wv} by following the QFT-like 3rd quantization paradigm together and using an additional `locality' assumption that restricts attention to an abelian sub-algebra, though this assumption can then be questioned as in
\cite{Giddings:1988wv}.   See also further comments in \cref{sec:Disc}.} see also \cite{Balasubramanian:2020jhl,Gesteau:2020wrk} and the Lorentz-signature discussion of baby universes in \cite{Marolf:2020rpm}.

Indeed, a key feature of the worldline formulation of  QFT \cite{Feynman:1951gn} is that the boundary conditions on worldline path integrals are associated with the arguments of correlation functions, and thus with the non-abelian algebra of quantum fields (see e.g., \cite{Strassler:1992zr,Schubert:2001he} for applications).  But the arguments of \cite{Marolf:2020xie} for baby-universe superselection sectors imply that boundary conditions of quantum gravity path integrals correspond to the arguments of correlation functions associated with an abelian algebra of operators that can be simultaneously diagonalized on the baby universe Hilbert space $\hbu$. This apparent tension was highlighted in \cite{Anous:2020lka}. To resolve this, below we focus on carefully identifying the steps in the QFT-like constructions that deviate from the framework defined in \cite{Marolf:2020xie}.  In particular, we will see that this issue is not related to any choice of spacetime signature, as both Lorentz- and Euclidean-signature constructions can in principle lead to either sort of algebra.  Instead, the critical issue is whether the inner product on the quantum gravity Hilbert space is constructed from an adjoint (or $\CPT$ conjugation) operation that leaves the set of allowed boundary conditions invariant.

We will proceed by example, exploring a series of  constructions one might use to relate path integral amplitudes to some quantum gravity inner product in various simple models.  Our main goal is to illustrate some key places where choices must be made, and where QFT-like approaches deviate from the framework of \cite{Marolf:2020xie}.  But this is only one step in analyzing the treatment of quantum gravity path integrals.  We will thus not concern ourselves with making the models particularly realistic.  In particular,  we will mostly study models which do not allow universes to split and join, so that our path integrals reduce to a collection of cylinders.  Though it is important to investigate such splitting and joining interactions in detail in the future, at least in perturbation theory, it is clear that adding such interactions to our models should cannot change any qualitative conclusions.

In fact, we will consider models of quantum gravity in which spacetime is one-dimensional (so that the above cylinders degenerate to become just line segments). Such models may thus be called ``worldline theories''.  This choice was made for simplicity and also for ease of comparison with QFT.  Since there is no concept of spatial boundary for one-dimensional Lorentzian spacetimes, our models are most analogous to studies of closed universes in higher dimensions.  In particular, we encourage the reader to think of the quantum gravity Hilbert spaces described below as analogues of the `baby universe sectors' of higher-dimensional models discussed in \cite{Coleman:1988cy,Giddings:1988cx,Giddings:1988wv,Polchinski:1994zs,Marolf:2020xie}.  We will thus often refer to them as baby universe Hilbert spaces below.  Comments on higher dimensional cases are interspersed throughout the text, but a full treatment of higher dimensional cases may require additional inputs beyond those discussed here.

\begin{table}[]
\centering
\begin{tabular}{l|c|c|c}
Choices & ESFTs  \cref{sec:EST}&  QFT-like \cref{sec:EQFT,sec:LQFT} &  GATs \cref{sec:GAT} \\
 \hline \hline
1) Proper Time (Lapse) Range ($\mathbb{R}^\pm$, $\mathbb{R}$) & $\mathbb{R}^+$  & $\mathbb{R}^+$ & $\mathbb{R}$   \\
2) Spacetime (Worldline) Signature (E,L) &  E & E/L & L   \\
3) Restricted Boundary Conditions & No & Yes & No  \\
4) $\CPT$ requires extra $\mathbb{Z}_2$? & No & For E target & No   \\
5) Target Space Signature  &  E & E/L & any but E \\
 \end{tabular}
\caption{The choices explored below associated with transforming a worldline path integral into a candidate quantum gravity Hilbert space, and the options chosen to define so-called Group Averaged Theories (GATs), QFT-like theories, and Euclidean Statistical Field Theories (ESFTs). E and L denote Euclidean and Lorentzian signatures respectively. Other terminology will be explained in the sections below. }
\label{tab:choices}
\end{table}

The set of options we explore is enumerated in  \cref{tab:choices}, though we defer a full explanation of the terms used there to the sections below.  We make no claim that this list is exhaustive.  In particular, we primarily consider unoriented spacetimes, only occasionally commenting on the possibility of including an orientation (e.g., in parallel with the treatment of \cite{Stanford:2019vob} for Jackiw-Teitelboim gravity).

We furthermore make no claim that any of the options described below correspond precisely to the way that specific models were studied in \cite{Marolf:2020xie} or in
\cite{Saad:2018bqo,Saad:2019lba,Blommaert:2019wfy,Saad:2019pqd,Penington:2019kki,Blommaert:2020seb}.  We thus defer any discussion of the detailed connection between those references and the approaches below to the discussion in \cref{sec:Disc}.

We begin in \cref{sec:reviewSS} below by reviewing the argument from \cite{Marolf:2020xie} that quantum gravity quantities associated with asymptotic boundaries defines an algebra of simultaneously-diagonalizable operators on a so-called `baby universe' Hilbert space.  This also provides an opportunity to provide an overview of the main structures required to define a quantum gravity Hilbert space from a gravitational path integral.  Some further preliminaries for one-dimensional gravity theories are then described in \cref{sec:general}.
We then proceed by examining various types of constructions in turn in  \cref{sec:EST,sec:EQFT,sec:LQFT,sec:GAT}, stepping through the ingredients and the choices to be made in each case.  For each type of construction, we focus on simple examples and indicate various potential generalizations without dwelling on the details. We conclude in \cref{sec:Disc} with a final discussion emphasizing open issues and the relation of our models to higher dimensional quantum gravity.

%~~~~~~~~~~~~~~~~~~~~~~~~~~~~~~~~~~~~~~~~~~~~~~~
\section{Boundary observables and Baby Universes}
\label{sec:reviewSS}
%~~~~~~~~~~~~~~~~~~~~~~~~~~~~~~~~~~~~~~~~~~~~~~

To set the stage for our discussion we begin with a recapitulating the essence of the argument from  \cite{Marolf:2020xie}. The  path integral for any quantum system defines a map from a set of boundary conditions to numbers, the amplitudes. The amplitudes depend on a prescription for the dynamics (the set of configurations to be summed over in the path integral and the corresponding weights, typically specified by an action), as well as the allowed set of boundary conditions.  We may superpose boundary conditions, turning the set of allowed boundary conditions into a vector space on which the map to amplitudes acts linearly.

\begin{figure}
\centering
\begin{tikzpicture}[scale=1]
\coordinate  (b) at (7,-1);
\coordinate (k) at (7,1);
\coordinate (a) at (0,0);
\draw[ball color=red!20, fill opacity=0.2,thick,black] (a) circle (2cm);
\draw[thick, ball color= green!30,fill opacity=0.4] (a) ellipse (2cm and 0.3cm);
\draw[ball color=red!20, fill opacity=0.2, thick,black] ($(k)+(2,0)$) arc (0:180:2);
\draw[thick,ball color= green!30,fill opacity=0.65, black] (k) ellipse (2cm and 0.3cm);
\draw[ball color=red!20,fill opacity=0.2, thick,black] ($(b)+(2,0)$) arc (0:-180:2);
\draw[thick, ball color= green!30,fill opacity=0.45, black] (b) ellipse (2cm and 0.3cm);
\node at ($(k)$) [below=8pt] {$\bra{\Phi}$};
\node at ($(b)$) [above=8pt] {$\ket{\Phi}$};
\node at ($(a) +(2,0)$) [right] {$\braket{\Phi}$};
\end{tikzpicture}
\caption{Slicing open a quantum amplitude to reveal the bra and ket components. }
\label{fig:slicing}
\end{figure}
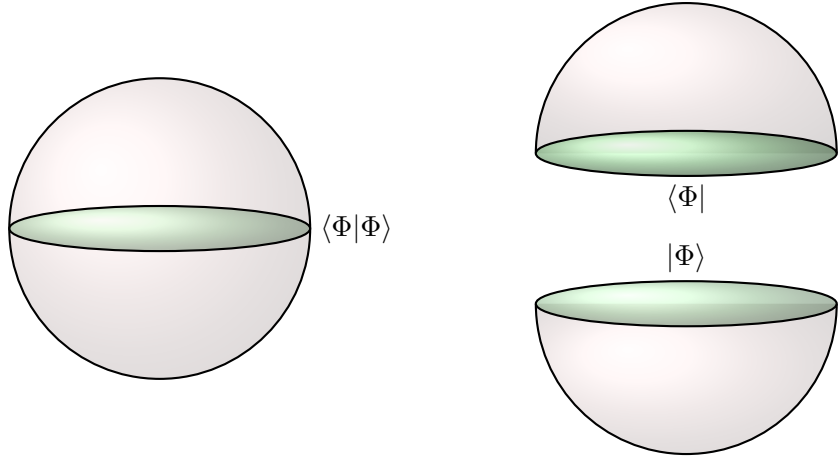

Given the quantum path integral, we can extract from it a Hilbert space by cutting it open along a codimension-1 slice. This follows from the convolutional property of path integrals; each cut corresponds to a resolution of the identity. The two halves of the path integral produced by the cut each use the boundary conditions appropriate to that part of the path integral to construct a state in the Hilbert space or in its dual; i.e., a ket-vector or a bra-vector.  The full path integral constructed by sewing them back together computes then the inner product between these bra- and ket- states, see \cref{fig:slicing}. In other words, we split the boundary conditions into `bra' and `ket' pieces (typically corresponding to future and past, respectively, when our cut is at some fixed time in QM or QFT), and the amplitudes define a bilinear product between these pieces.  To obtain a Hilbert space (with an inner product between two ket-vectors, say) we must have an anti-linear map turning a ket boundary condition for the path integral into a bra boundary condition, which squares to the identity.  We may think of this as prescribing the action of  a \CPT map on the space of allowed boundary conditions for the path integral.

Now, if the quantum gravity path integral is a sum over all topologies, then it naturally allows topologies with an arbitrary number of boundaries.  As a result, if $\BC$ is the space of allowed boundary conditions at a single boundary, then
any list $b_1,\ldots b_n$ of elements $b_i \in \BC$ (of any length $n$) will define an allowed boundary condition for our quantum gravity theory.  Furthermore, the path integral is to be computed by summing over \emph{all} spacetimes with boundaries matching $b_1,\ldots , b_n$.  Thus the ordering of the boundary conditions plays no role, and two lists that differ by a permutation should be viewed as defining the same boundary conditions.    Using $\expval{ b_1,\ldots, b_n}$ to denote the path integral with boundary conditions $b_1,\ldots, b_n$, we may then write
\begin{equation}
\label{eq:BUperm}
\expval{ b_1, b_2, \ldots, b_n }
= \expval{ b_{\sigma(1)}, b_{\sigma(2)}, \ldots, b_{\sigma(n)} } \,,
\end{equation}
where $\sigma \in S_n$ is a permutation of the boundary conditions. This turns the vector space of boundary conditions into an abelian algebra, with the product defined by disjoint union of boundaries.

As noted above, we should be able to cut open such a path integral to define a state. Due to the presence of boundaries, there are various ways in which we could introduce such a cut.  For simplicity let us introduce a cut that does not intersect any of the existing boundaries $b_1, b_2, \ldots, b_n$, but merely partitions them into two disjoint subsets.  Since each piece of the resulting path integral should define a state on the cut, there should be states $\ket{ a_1,\ldots , a_m}$ associated with arbitrary lists of boundary conditions, where the symmetry \eqref{eq:BUperm} means that we must identify states that differ only by the ordering of the boundary conditions $a_1,\ldots ,a_m $.  And since the above cuts  are closed surfaces, we should think of these states as describing closed universes without boundary.  As a result, it is traditional to call this the Hilbert space $\hbu$ of `baby universes,' with the idea that such closed universes may have somehow been `produced' by some larger (infinite) `parent universe' having a non-trivial asymptotic boundary.

For each allowed boundary condition $b \in \BC$, it is then natural to define an operator $\hat b$ on $\hbu$ that simply inserts an additional boundary with the stated boundary condition; i.e.,
\begin{equation}
\hat b\ket{ a_1,\ldots, a_m } =  \ket{b, a_1,\ldots, a_m }.
\end{equation}
Since the ordering of the boundary conditions is unimportant, it is manifest that any two such operators commute:
\begin{equation}\label{eq:commute}
\hat b_1 \hat b_2  \ket{a_1,\ldots , a_m} =  \ket{b_1, b_2, a_1,\ldots , a_m}=  \ket{b_2, b_1, a_1,\ldots , a_m}
= \hat b_2 \hat b_1  \ket{a_1,\ldots , a_m}.
\end{equation}
Finally, there is one special state $\ket{\HH}$ (the Hartle-Hawking no-boundary state) which corresponds to the absence of boundaries, $m=0$. All the states of $\hbu$ are then generated by the action of the algebra of boundary-inserting operators $\hat{b}$ acting on $\ket{\HH}$,
\begin{equation}
	\ket{b_1,\ldots, b_n} = \hat{b}_1 \cdots  \hat{b}_n \ket{\HH},
\end{equation}
and linear combinations.

As noted above, this is not quite enough to define the inner product on $\hbu$, since in addition we must choose an anti-linear `$\CPT$' operation acting on boundary conditions. Indeed, we will see below that a single set of amplitudes may be associated with several different Hilbert spaces, by making a different choice of this conjugation operation. This choice is equivalent to defining the adjoint of the boundary-inserting operators $\hat{b}$ (and the conjugate to the no-boundary state). We then have
\begin{equation} \label{eq:QGIP1}
	\begin{aligned}
		\braket{ a_1,\ldots , a_m }{b_1,\ldots , b_n }
	&=
		\mel{\HH}{ \,\hat{a}_1^\dag \cdots \hat{a}_m^\dag \,\hat{b}_1 \cdots  \hat{b}_n \,}{\HH} \\
	&=
		\expval{ a_1^\dag, \ldots , a_m^\dag, b_1,\ldots , b_n }\,,
	\end{aligned}
\end{equation}
where the second line is just a path integral amplitude written in the same notation as in \eqref{eq:BUperm}, and where we have assumed that the application of \CPT to a list of boundary conditions $a_1,\ldots a_m$ is given by applying \CPT to the individual members of the list. This means that for any $b\in\BC$, $\hat{b}^\dag$ acts by inserting some boundary $b^\dag\in\BC$ (reusing the adjoint notation from the operator interpretation), so $\dag$ acts on the space of connected boundaries and extends to multiple boundaries in the simplest possible manner.\footnote{In particular, this means that the no-boundary condition is left invariant, so the norm of $\ket{\HH}$ is given by the path integral over closed spacetimes with no boundary whatsoever. We may also choose to normalize $\ket{\HH}$, which means defining the amplitudes to include a denominator of the path integral over closed spacetimes. Equivalently, we can integrate only over spacetimes without closed components, in the same way that vacuum diagrams are removed by normalization in QFT.\label{foot:HHnorm}}

This defines a sesquilinear product on the space of boundaries. For this to give a sensible Hilbert space, we must require that it is positive semi-definite; that is, the norm of any state thus computed is nonnegative. Under that assumption, we can define $\hbu$ as the completion of the span of states $\left| b_1,\ldots, b_n \right\rangle$ (i.e., the completion of polynomials in boundary conditions $b_i$) with the given inner product.\footnote{Defining $\hbu$ from the amplitudes of the algebra $\BC$ in the Hartle-Hawking state in this way is closely analogous to the GNS construction \cite{Gelfand:1943imb,Segal:1947irr} (see also \cite{Gesteau:2020wrk,Anous:2020lka}), but not technically identical since we do not a priori have a norm on the space of operators.} Since the inner product is required only to be positive \emph{semi}-definite, nontrivial linear combinations of the states $\left| b_1,\ldots, b_n \right\rangle$ can be `null': they have zero norm, and hence in the completion $\hbu$ are equal to the zero state. One can informally say that $\hbu$ is constructed as a quotient by such null states (though it is not technically necessary to invoke such a quotient to define the completion).

 However, there are two more issues that we should consider. The first is to show that our boundary-inserting operators $\hat{b}$ are truly well-defined on $\hbu$. The potential issue here arises due to the above quotient by null states, since $\hat{b}$ is a well-defined operator on $\hbu$ only if it preserves the space of null states. But this is straightforward to show from \eqref{eq:QGIP1}. The key observation is that $\hat{b}$ acting to the right is equivalent to $\hat{b}^\dag$ acting to the left, and have assumed that $\hat{b}^\dag$ acts by adding some boundary $b^\dag$:
 \begin{equation}
 	\mel{ a_1,\ldots , a_m }{\hat{b}}{b_1,\ldots , b_n}
 		=  \mel{ a_1,\ldots , a_m }{(\hat{b}^\dag)^\dag}{b_1,\ldots , b_n}
 		= \braket{ a_1,\ldots , a_m, b^\dag  }{b_1,\ldots , b_n}.
 \end{equation}
As a result, for any null state $\ket{N}$ and any boundaries $a_1,\ldots , a_m$ we have
\begin{equation}
\label{eq:null}
\mel{ a_1,\ldots , a_m }{ \hat b }{N } =
\braket{  a_1,\ldots , a_m, b^\dag}{N } =0,
\end{equation}
where the last equality holds because $\ket{N}$ is null. This means that the overlap of $\hat{b} \ket{N}$ with any state is zero, so $\hat{b} \ket{N}$ is also null, and thus $\hat b$ preserves the null space as desired.

Secondly, we would also like to show that the $\hat{b}$ can be simultaneously diagonalized.  In general these operators are not Hermitian, but they are \emph{normal}, meaning that $\hat{b}$ commutes with its adjoint $\hat b^\dagger$, $\commutator{\hat{b}}{\hat{b}^\dag}=0$. This follows from the fact that $\hat{b}^\dagger$ also acts by inserting a boundary $b^\dag$, so we can apply \eqref{eq:commute} with $b_1=b$ and $b_2=b^\dag$. This means that we can apply the spectral theorem, so $\hat{b}$ is diagonalizable. In fact, all of the operators $\hat{b}$ and $\hat{b}^\dagger$ commute, and hence all $\hat{b}$ can be simultaneously diagonalized as desired. The baby universe Hilbert space $\hbu$ has a basis of simultaneous eigenvectors $\ket{\alpha}$ for all boundary-inserting operators $\hat{b}$, labeled by some (continuous or discrete) parameters $\alpha$:
\begin{equation}\label{eq:alphastates}
	\hat{b} \ket{\alpha} = b_\alpha \ket{\alpha}
\end{equation}
for some $b_\alpha\in\CC$, for all $b\in \BC$. These $\alpha$-states give superselection sectors for the commutative algebra generated by boundary-inserting operators.

It is clear that the above argument is very general.  The key point is simply that quantum gravity inner products are given by path integral amplitudes as in \eqref{eq:QGIP1} for some set of single-boundary boundary conditions $\BC$ that is invariant under the action of $\dag$.  While this appears to us to be a natural condition to impose on theories of quantum gravity, as we review below it is certainly \emph{not} the case for the construction of QFT from worldline path integrals.  This is illustrated by the examples below in which we also discuss certain other choices that must be made to define a Hilbert space from quantum-gravity-like path integrals.

%~~~~~~~~~~~~~~~~~~~~~~~~~~~~~~~~~~~~~~~~~~~~~~~
\section{One-dimensional Theories of Gravity}
\label{sec:general}
%~~~~~~~~~~~~~~~~~~~~~~~~~~~~~~~~~~~~~~~~~~~~~~

We now describe a general framework that forms the backbone of our one-dimensional quantum gravity models.  Thinking of a gravitational theory as a path integral over spacetimes, we first describe the amplitudes resulting from the sum over one-dimensional manifolds. For the sake of simplicity we will focus on the non-interacting limit (the free theory), where our worldlines do not intersect with each other.  We then discuss the choice of  `matter' degrees of freedom which lives on these spacetimes, giving the prominent example of a minisuperspace model obtained as a dimensional reduction of a theory in higher dimensions.

These ingredients will however not suffice to characterize our quantum gravitational theories completely: there will be additional discrete choices that need to be made in order to fully specify the model as summarized in \cref{tab:choices}. We will  explore these choices in detail  subsequently in  \cref{sec:EST,sec:EQFT,sec:GAT,sec:LQFT}.

%~~~~~~~~~~~~~~~~~~~~~~~~~~~~~~~~~~~~~~~~~~~~~~~
\subsection{Amplitudes from the sum over one-dimensional spacetimes}
\label{sec:1damps}
%~~~~~~~~~~~~~~~~~~~~~~~~~~~~~~~~~~~~~~~~~~~~~~

For given boundary conditions, our one-dimensional gravity amplitudes will be defined by an integral over the compatible one-dimensional manifolds, perhaps with some `matter' quantum mechanics living on those manifolds.
Fortunately one-dimensional manifolds and metrics are simple: we have only intervals or circles, parameterized by their total
proper length $T>0$. Here we treat the real line as the $T\rightarrow \infty$
limit of an interval.

Since we are interested in the dependence on boundary conditions, we may ignore the circles (which in any case only contribute an overall normalization; see \cref{foot:HHnorm}) and restrict to intervals. For the most part  we take the spacetimes we sum over to be simply a union of intervals, though we will take occasion to comment on the generalization to graphs, where we allow several intervals to be sewn together at their boundaries. Throughout our discussion, we will use quantum gravity terminology so that the one-dimensional manifold is a `spacetime,' and a single such manifold represents a `universe' (and not a `particle').  It will, however, sometimes be convenient to also refer to the spacetime as the `worldline' of the universe.

The boundary of a one-dimensional spacetime is a (zero-dimensional) collection of points. As above, we use $\BC$ to denote the set of allowed boundary conditions at any single such point.\footnote{If we were to sum over oriented spacetimes, we should also assign an orientation to the boundary, which means a choice of sign for each point.} Note that in one-dimensional gravity theories these will be conditions on the `matter' fields alone as the space of zero-dimensional metrics is trivial. For example, when one defines a one-dimensional quantum gravity model by Kaluza-Klein reduction of a higher-dimensional theory, most of the features the gravitational theory in fact become part of the Kaluza-Klein matter sector, with only an overall notion of proper time left to be treated as one-dimensional gravity.

Following the discussion of \cref{sec:reviewSS}, a general quantum gravity boundary condition is an unordered list of elements of $\BC$. The associated quantum gravity amplitude is to be defined by summing over all one-dimensional manifolds compatible with the boundary conditions.  Since for now our one-dimensional manifolds are collections of $n$ intervals, in the absence of inter-universe interactions, the number of boundaries must be an even number $2n$ for the amplitude to be nonzero. The topology of spacetime is specified by assigning the boundaries into $n$ pairs, where an interval connects the points in each pair.\footnote{For oriented spacetimes, the endpoints of an interval must have opposite orientation. The spacetime topologies are equivalent to maps from negatively oriented points to positively oriented points, which must be equal in number.} Each resulting pair of boundary conditions then specifies a single-universe amplitude computed by integrating over metrics and corresponding matter fields on the interval. The full quantum gravity amplitude is then formed by multiplying together the $n$ single-universe amplitudes defined by each pairing and then summing over pairings.  For $b_1,\ldots, b_n \in \BC$, the associated quantum gravity amplitude may thus be written
\begin{equation}
\label{eq:QGA}
 \expval{ b_1,\ldots,b_{2n}} = \sum_{\substack{\text{pairings of} \\\{b_1,\ldots,b_{2n}\}}}  \  \prod_{\substack{ \mathrm{pairs} \\ \{b_i,b_j\}} } \ASU(b_i,b_j)\ ,
\end{equation}
where $\ASU(b_i,b_j) = \langle b_i b_j\rangle$ is the `single-universe amplitude' associated with integrating over the parameters of a single interval with the boundary conditions specified by the pair $b_i,b_j$. As in \cite{Marolf:2020xie}, we treat all boundaries as being distinguishable so that there are no additional symmetry factors in \eqref{eq:QGA}.  Our non-interacting universe assumptions implies \eqref{eq:QGA} has the form of Wick contractions, so that in simple cases the quantum gravity amplitudes can be written as Gaussian integrals over the space of boundary conditions with a covariance matrix specified by the single-universe amplitudes $\ASU$, and the baby universe Hilbert space will be a Fock space built on a single-universe Hilbert space.\footnote{ Should we include interactions between worldlines we would have to consider summing over graphs with intermediate splitting and joining of worldlines.  This can be accounted for perturbatively by including in \eqref{eq:QGA} the sums over appropriate collection of graphs with fixed boundaries. In this case the quantum gravity Hilbert space is no longer the Fock space over a single-universe Hilbert space.% It would be useful to construct toy models with such interactions to better  understand  the structure of the Hilbert space.   
}

Note in particular that \eqref{eq:QGA} is invariant under arbitrary permutations of the single-universe boundary conditions
$b_1,\ldots,b_n$. This may remind the reader of bosonic quantum field theory, and thus raise questions of whether other possibilities might be allowed as well.  One might also ask about further modifications of \eqref{eq:QGA}. While such questions may be of interest, we will not explore them below.  Instead, we take the structure embodied in \eqref{eq:QGA} as given and consider additional choices that must be made in order to both define the single-universe amplitudes used in \eqref{eq:QGA} (discussed below) and in order to construct the quantum gravity Hilbert space from the above quantum gravity amplitudes (discussed in \cref{sec:EST,sec:GAT}).

To construct the full amplitudes it remains to define the single-universe amplitudes $\ASU$, which requires two ingredients. The first is a definition of the integral over spacetime (worldline) metrics on each interval.  In practice, this will involve specifying whether the signature is Lorentzian or Euclidean and choosing a range of integration for the length (or proper time) $T$ of each interval. The second is the specification of the `matter' model and its associated dynamics and boundary conditions.

Explicitly, we can write $\ASU$ with an integral over spacetimes (labeled by their length $T$) and over matter fields, labeled $x$:
\begin{equation}\label{eq:asudef}
\ASU(b_1,b_2) = \int_D dT \int_{b_1}^{b_2} \mathcal{D}x \, e^{\eta\, S_\mathrm{matter}[x;T]} \,,
\end{equation}	
where $\eta = +i$ and $\eta = -1$ for Lorentzian and Euclidean worldlines, respectively, and $b_1,b_2$ indicate some choice of boundary conditions for matter fields $x$ at each end of the worldline. We may want to add a `gravitational action' depending only on metrics, but the only local possibility in one dimension is a `cosmological constant' proportional to $T$, which we have chosen to absorb as a constant shift of the matter Lagrangian. The flat measure $dT$ over metrics is fixed by locality \cite{Cohen:1985sm} (and may also be obtained by starting with a general metric and gauge-fixing one-dimensional diffeomorphisms by the Faddeev-Popov method).\footnote{Non-trivial measures for the proper-time integration have been previously considered, see eg., \cite{Abel:2019ufz,Abel:2019zou}.} For further discussion of the measure we refer the reader to \cite{DeBoer:1995hv,Bastianelli:2006rx,Bastianelli:2013tsa,Edwards:2019eby}. Finally, we must choose a range of integration $D$ for the lapse or proper time $T$, with the only choices respecting locality being $\RR$, $\RR^+$ or $\RR^-$.

To define the amplitudes, the main choices open to us are explicit in \eqref{eq:asudef}:
\begin{itemize}
	\item The worldline signature $\eta$.
	\item The range of proper time $D$.
	\item The matter dynamics (fields $x$ and action $S_\mathrm{matter}$).
	\item The allowed boundary conditions $b$ for matter.
\end{itemize}
As we will see in the course of our discussion, it is possible to construct examples that differ from the worldline formalism of QFTs by our choice of the domain $D$.

In addition, to define the quantum gravity Hilbert space we must specify the $\CPT$ operation $\dag$. It is important to note that $\dag$ is not in general determined by the choices above, and that it plays a critical role in the theory.  In particular, the quantum gravity theories of sections \ref{sec:EST} and \ref{sec:EQFT} are distinguished only by the choice of $\dag$.

 We will outline the class of matter theories we study below, and the remainder of the paper will be organized by various permutations of the other choices, and devoted to discussing their consequences for the resulting baby universe Hilbert space.

There are some additional choices that are less important for our purposes, and will only be mentioned parenthetically. First, we could choose our spacetimes to carry an orientation; we will mostly concentrate on unoriented worldlines, which requires restricting to matter theories with time-reversal symmetry (c.f., the discussion of JT gravity in two dimensions \cite{Stanford:2019vob}). Likewise, we could choose to have worldline supersymmetry, which for instance provides an example (once we make some of our discrete choices)  of a topological sigma model \cite{Witten:1982im} (see \cite{Birmingham:1991ty} for the classic review of these developments). Finally, a more radical generalization  is to sum not only over disjoint unions of intervals but general graphs, which we will have occasion to comment on in various cases.

%~~~~~~~~~~~~~~~~~~~~~~~~~~~~~~~~~~~~~~~~~~~~~~~
\subsection{The matter theory}
\label{sec:matterchoice}
%~~~~~~~~~~~~~~~~~~~~~~~~~~~~~~~~~~~~~~~~~~~~~~

The main choice which determines the amplitudes $\ASU$ will be the matter theory. We can describe this either by a path integral, specifying matter fields and Lagrangian (as we have done above), or by a Hilbert space of matter states and a Hamiltonian $H$.   In the Hamiltonian formulation, the contribution to $\ASU$ from an interval with proper length $T$ will be given by matrix elements of $e^{-iHT}$ or $e^{-HT}$ for Lorentzian or Euclidean spacetimes respectively, between states determined by the boundary conditions.

In describing each class of examples below, we will focus mostly on a matter theory taking the form of a sigma-model, so that the field content is a map from the one-dimensional spacetime to some target space $\target$.  We take this target space to be equipped with a metric $g_{\mu\nu}(x)$,  and thus with a Laplacian (or wave operator) $\nabla^2$. The matter action takes the form
\begin{equation}\label{eq:smatter}
S_\mathrm{matter} = \int d\tau \left[ \tfrac{1}{2} g_{\mu\nu} \dv{x^\mu}{\tau}\, \dv{x^\nu}{\tau}  -  U(x)\right] ,
\end{equation}	
allowing for the possibility of a potential $U(x)$ on $\target$. Equivalently, the Hamiltonian is
\begin{equation}\label{eq:Hmatter}
H = -\tfrac{1}{2}\nabla^2 + U(x) \,,
\end{equation}	
acting on the Hilbert space $L^2(\target)$.

The most general possible boundary condition is to give a wavefunction valued in $\target$, that is a state in the matter Hilbert space $L^2(\target)$. This is a linear combination of boundary conditions that fix the fields to take a specific value $x\in \target$ at the corresponding endpoint of spacetime, corresponding to delta-function wavefunctions $|x\rangle_\mathrm{matter}$, where we use the matter label to avoid confusion with $\hbu$. Thus, we will label our boundary conditions as points $x$, so the amplitudes will be denoted $\langle x_1,\ldots,x_n\rangle$ for lists of points in $\target$.

Models of this kind naturally arise in the so-called mini-superspace truncation of gravity \cite{Misner:1972ab,Misner:1973zz} in which one restricts a higher-dimensional model of gravity to e.g., homogeneous spacetimes. At this stage we will not restrict the signature of $\target$, so in particular the Hamiltonian may not be bounded below.\footnote{ We will take the metric to have mostly positive signature for Lorentzian targets.}

We will however  require two properties that are \emph{not} obviously natural in the most straightforward construction of such minisuperspace models.  In particular, we will take $\target$ to be geodesically complete, so that  $\nabla^2$  defines an essentially self-adjoint operator on $L^2(\target)$. If $\target$ is Lorentzian then $\nabla^2$ will have a continuous spectrum, while for Euclidean target one may end up with a discrete spectrum if $\target$ is compact.

%~~~~~~~~~~~~~~~~~~~~~~~~~~~~~~~~~~~~~~~~~~~~~~~
\subsection{Mini-superspace models}
\label{sec:minisuper}
%~~~~~~~~~~~~~~~~~~~~~~~~~~~~~~~~~~~~~~~~~~~~~~

To provide additional context for this treatment of the one-dimensional gravitational sector, let us briefly discuss an example of a minisuperspace model (see e.g., \cite{Ryan:1975jw,Kodama:1997tk}). The dynamics of spatially homogeneous $3+1$ Lorentz signature Einstein-Hilbert gravity on a 3-torus with vanishing cosmological constant, the \emph{Bianchi I model},  is closely related to that of a free massless particle in 2+1 Minkowski space (say, with inertial coordinates $x^0, x^1, x^2$), see \cite{Misner:1974qy}.  In particular,  with $Y^i$ for $i=1,2,3$ being coordinates on the spatial 3-torus,  consider the cosmological geometry with metric:
 \begin{equation}
\gamma_{ab}\, dX^a\, dX^b = - N_0(t)^2\,  dt^2 + e^{2 x^0(t)}\, (e^{2X(t)})_{ij} \,dY^i dY^j
 \end{equation}
Here $X(t)$ is a diagonal matrix after fixing the diffeomorphisms symmetries sans time reparameterization; specifically we take
$X(t) = \operatorname{diag}\{x^1(t) +\sqrt{3}\, x^2(t), x^1(t) - \sqrt{3}\, x^2(t), -2x^1(t)\}$ to describe the anisotropies. The overall scale-factor of the torus is given by $e^{x^0}$.  Using the standard lapse function $N_0$ that measures proper time and momenta $p_0, p_1,p_2$ conjugate to $x^0, x^1, x^2$, viz., $p^\mu = \dv{x^\mu}{t}$, the Einstein-Hilbert Lagrangian  may be written as (nb: $\mu\in \{0,1,2\}$)
 \begin{equation}
 \sqrt{-\gamma}\ {}^\gamma R  = \dot{x}^\mu p_\mu  - N_0\,H_0\,.
 \end{equation}
Here we introduced, $H_0$, the Hamiltonian constraint, given by
 \begin{equation}
H_0  = \frac{e^{-3x_0} }{24} (-p_0^2 + p_1^2 + p_2^2) \,.
 \end{equation}
Now,  $H_0$ vanishes on-shell   due to the equation of motion obtained by varying $N_0$. Owing to the prefactor $e^{-3x_0}$, the constraint $H_0$ tends to generate evolution that reaches a cosmological singularity at $x_0 = -\infty$.
However,  using a rescaled lapse $N = \frac{1}{12} \, N_0\,  e^{-3x_0}$  and the associated rescaled constraint
 \begin{equation}
 \label{eq:rescaledB1}
H  =\frac{1}{2}\left( -p_0^2 + p_1^2 + p_2^2\right) ,
 \end{equation}
we may recast the dynamics in the advertised form of a standard massless particle in 2+1 Minkowski space $\mathbb{R}^{2,1}$.

In particular, reduction to 0+1 dimensions yields a `matter' theory defined by \eqref{eq:rescaledB1}, which we may think of a sigma-model with target space $\target = \mathbb{R}^{2,1}$.  The gravitational sector of the reduced theory naturally has Lorentz signature, and we {\it define} the `proper time' $T$ of the reduced theory by $dT = N dt$.  Note that this differs from the natural notion of proper time $N_0dt$ associated with the 3+1 geometry, though it corresponds to the usual notion for a massless particle on $\mathbb{R}^{2,1}$.

We offer this model as an illustration of way to obtain Lorentz signature target spaces for the worldline theory. Note that from the higher dimensional point of view we are only attempting to keep track of a subset of gravitational degrees of freedom in the minisuperspace approximation. One furthermore is also holding the higher dimensional topology fixed. Given the natural  tendency of gravitational dynamics to lead to cosmological singularities, ${\cal M}_{\text{target}}$ is typically not geodesically complete when the metric on it is defined by the constraint associated with evolution in proper time.  For now we note only that this difficulty can typically be circumvented by using a rescaled `conformal time' dynamics near the singularity (which amounts to the use of singularity-avoiding coordinates), so that we can readily construct gravity-inspired models of the mathematical form described here.  But we will return to discuss the physics of this rescaling in \cref{sec:Disc}.

%~~~~~~~~~~~~~~~~~~~~~~~~~~~~~~~~~~~~~~~~~~~~~~~
\section{Euclidean Statistical Theories}
\label{sec:EST}
%~~~~~~~~~~~~~~~~~~~~~~~~~~~~~~~~~~~~~~~~~~~~~~

For our first two examples, discussed in this section and the next, we will choose the worldline quantum gravity theory to have Euclidean signature. While this is perhaps not as interesting as the Lorentzian theories discussed later from the point of view of higher-dimensional models of quantum gravity, we present it first as a clean and familiar setting to illustrate the impact of certain choices made to define the Hilbert space.

In fact, in both this section and in \cref{sec:EQFT} we will use precisely the same amplitudes, identified with correlation functions of a Euclidean field theory, but nonetheless construct two different Hilbert spaces. The choices we make here will lead to a baby universe Hilbert space $\hbu$ which is very natural if we interpret our field theory as a classical statistical model. A different set of choices (designed to make contact with a quantum field theory Hilbert space) will be discussed in \cref{sec:EQFT}.

%~~~~~~~~~~~~~~~~~~~~~~~~~~~~~~~~~~~~~~~~~~~~~~~
\subsection{The amplitudes}
\label{sec:ESTamp}
%~~~~~~~~~~~~~~~~~~~~~~~~~~~~~~~~~~~~~~~~~~~~~~

Taking our spacetimes to have Euclidean signature means that the matter amplitudes on an interval of length $T$ are given by the matrix elements  $\langle x | e^{-HT} | y \rangle_\mathrm{matter}$.   Our use of unoriented worldlines means that we have an $x \leftrightarrow y$ symmetry, so these matrix elements are symmetric and real. After integrating over one-dimensional metrics, the single-universe amplitudes $\ASU(x,y)$ that enter the quantum gravity amplitudes \eqref{eq:asudef} thus take the form
\begin{equation}
\label{eq:ESTASU}
\ASU(x,y) = \int_{\cal D} dT \, \mel{x}{e^{-HT}}{y}_\mathrm{matter}
\end{equation}
for some choice of integration domain ${\cal D}\subset {\mathbb R}$.  The three local options are just the positive reals ${\mathbb R}^+$, the negative reals ${\mathbb R}^-$, or the entire real line.  But for convergence, we require that the Hamiltonian $H$ is positive, and $\mathcal{D} = \mathbb{R}^+$ (or $H$ is negative, in which case we may replace $H\to -H$ w.l.o.g.). For a sigma-model as in \eqref{eq:Hmatter}, this means that the target space $\target$ must have Euclidean signature. With this choice, the resulting amplitudes are given by the matrix elements of the inverse of the Hamiltonian:
\begin{equation}
\label{eq:ESTASU2}
\ASU(x,y) = \int_{{\mathbb R}^+} dT \,\mel{x}{e^{-HT}}{y}_\mathrm{matter} = \mel{x}{\frac{1}{H}}{y}_\mathrm{matter}.
\end{equation}
In other words, $\ASU(x,y)$ is the Green's function for the Hamiltonian $H$ (with appropriate fall-off conditions if $\target$ is non-compact).

For general amplitudes, we recall that \eqref{eq:QGA} is computed from \eqref{eq:ESTASU2} from Wick contractions, and can thus be written in terms of an appropriate Gaussian path integral over a real scalar field $\Phi$ valued in $\target$:
\begin{equation}
\label{eq:ESTQGA}
\begin{gathered}
	\expval{ x_1,x_2,\ldots, x_n} =
	\mathcal{N} \int {\cal D}\Phi \, \ \Phi(x_1)  \cdots  \Phi(x_n)  \; e^{-I[\Phi]},\\
	\text{where}\quad I[\Phi] = \int_{\target}dx\bigg[\tfrac{1}{2}(\partial \Phi(x))^2+U(x) \Phi(x)^2\bigg],
\end{gathered}
\end{equation}
with $\mathcal{N}$ a normalization constant.  In other words, our quantum gravity amplitudes are the correlation functions of a free Euclidean field theory on $\target$, consisting of a scalar field $\Phi$ with action $I$.

%~~~~~~~~~~~~~~~~~~~~~~~~~~~~~~~~~~~~~~~~~~~~~~~
\subsection{The baby universe Hilbert space}
\label{sec:ESTBU}
%~~~~~~~~~~~~~~~~~~~~~~~~~~~~~~~~~~~~~~~~~~~~~~

As already emphasized, the amplitudes alone do not provide sufficient information to construct the baby universe Hilbert space $\hbu$. In particular, in order to define the inner product we must additionally choose a conjugation operation, `$\dag$', acting on boundary conditions. Here, we will make the most obvious choice, that $\dag$ acts trivially on the $x$ boundary conditions:
\begin{equation}
	x^\dag = x.
\end{equation}

From this, we can construct and interpret the baby universe Hilbert space. As noted earlier, since the general inner product is computed by Wick contractions, $\hbu$ is a Fock space built on the single-universe Hilbert space, spanned by $ \ket{x}$ for $x\in\target$. A general single universe state is a superposition $\int dx\, F_1(x) \ket{x}$ for some complex-valued function $F_1$ on $\target$. The inner product of two such states $F_1,G_1$ is constructed from the single-universe amplitude \eqref{eq:ESTASU2}, as
$\int dx dy \,G_1^*(x)\ASU(x,y)F_1(y)$. This is positive-definite, which can be seen by decomposing $F_1$ in a basis of eigenfunctions of $H$ and using positivity of the corresponding eigenvalues. In particular, there are no nontrivial null (zero-norm) states.

There is in fact a nicer characterization of the full Hilbert space $\hbu$, as a space of (complex-valued) functionals $F$ of our scalar field $\Phi$ taking values in $\target$. Formally, we may write a functional $F$ in terms of its Taylor expansion,
\begin{equation}
	F[\Phi] = F_0 + \int dx\, F_1(x)\Phi(x) + \tfrac{1}{2}\!\int dx_1 dx_2\, F_2(x_1,x_2)\Phi(x_1)\Phi(x_2) + \cdots,
\end{equation}
where $F_n$, the $n^\text{th}$ functional derivative of $F$, is a symmetric function $\target^n\to \CC$. In particular, the one-universe Hilbert space considered above corresponds to linear functionals, where all $F_n$ vanish for $n\neq 1$. We can then write any state of $\hbu$ in terms of a functional $F$, as
\begin{equation}\label{eq:functionalState}
	 \ket{F} = F_0 \ket{\HH} + \int dx\, F_1(x) \ket{x} + \tfrac{1}{2}\!\int dx_1 dx_2\, F_2(x_1,x_2)\ket{x_1,x_2} + \cdots,
\end{equation}
and the inner product of two such states is given by
\begin{equation}
	\braket{G}{H} = \mathcal{N} \int {\cal D}\Phi \, G^*[\Phi] \,F[\Phi] \; e^{-I[\Phi]}.
\end{equation}
This holds term-by-term in the Taylor expansion, since both Gaussian integrals and our quantum gravity amplitudes \eqref{eq:QGA} are computed by Wick contractions.

In this form, we may see that $\hbu$ has a simple interpretation if we take the path integral to define a classical statistical system, such as the continuum limit of an Ising-like model. The path integral $\int \mathcal{D}\Phi \,\cdots e^{-I[\Phi]}$ defines a probability distribution (perhaps a Boltzmann distribution where $I[\Phi]$ is $\beta$ times the energy of the given field configuration). The functionals $F$ are then the observables of such a model, which are random variables depending on the probability distribution. The inner product then gives us the covariance matrix of these random variables,
\begin{equation}
	\left\langle G \middle|F	 \right\rangle = \operatorname{Cov}(F,G).
\end{equation}
Thus, $\hbu$ is a standard construction in probability theory, the Hilbert space of random variables (of finite variance).

Moreover, it is now simple to interpret the superselection sectors (or $\alpha$-states) of $\hbu$: they are states where the field $\Phi$ takes a definite value, and the eigenvalues of boundary operators $\hat{x}$ are given by $\Phi(x)$. The action $e^{-I[\Phi]}$ gives the square of the overlap of (appropriately normalized) $\alpha$ states with the no-boundary state $\ket{\HH}$, so the probability distribution of superselection sectors is precisely identified with the original distribution defining the classical statistical theory.

%~~~~~~~~~~~~~~~~~~~~~~~~~~~~~~~~~~~~~~~~~~~~~~~
\subsection{Generalizations}
%~~~~~~~~~~~~~~~~~~~~~~~~~~~~~~~~~~~~~~~~~~~~~~
\label{sec:ESTgeneralizations}

One can exemplify the construction above with many familiar examples where $\target$ is compact. We could just as well also choose a non-compact target geometry.  For definiteness, consider the Wick-rotation of the  Bianchi I model  introduced in \cref{sec:matterchoice} -- the replacement $x^0 \rightarrow i x_3$  yields\footnote{Here we again use the `rescaled lapse' $N = e^{-3x_0}N_0$.}
\begin{equation}
\label{ESTH}
H = p_1^2 + p_2^2 + p_3^2
\end{equation}
in terms of the Euclidean target space $\BC = {\cal M}_{\text{target}} = {\mathbb R}^3$.

From \eqref{eq:ESTASU2}, it is clear that we may generalize this approach to allow any matter theory with a positive-definite Hamiltonian,  a sufficiently well-defined resolvent operator, and time-reflection symmetry.  Notably, this excludes the case of Lorentz-signature target spaces as naturally occur in quantum gravity models.  However, if there is an appropriate ${\mathbb{Z}}_2$ symmetry as above, one can Wick-rotate such models to Euclidean signature and then apply the above approach.  As before, the time-reflection symmetry is required due to our choice to sum over unoriented spacetimes.

Considering instead oriented spacetimes removes this requirement (for example, allowing us to include a background magnetic field for our particle), both giving a time orientation for the dynamics and an orientation (a sign) for boundary conditions. In that case, we denote a Dirichlet boundary condition with positive orientation by $\Phi(x)$, and one with negative orientation by $\bar{\Phi}(x)$. Any spacetime is a union of intervals connecting a $\Phi$ boundary with a $\bar{\Phi}$ boundary, so the amplitudes are
\begin{equation}
\label{eq:ESTQGAC}
\begin{split}
	\Big\langle \Phi(x_1) \cdots \Phi(x_n) &\,\bar{\Phi}(y_1)\cdots \bar{\Phi}(y_m)\Big\rangle
	= \\
& {\cal N} \int {\cal D}\Phi {\cal D}\bar \Phi\ \Phi(x_1)\cdots \Phi(x_n) \,\bar{\Phi}(y_1)\cdots \bar{\Phi}(y_m) \,
	 e^{-\frac{1}{2}\, \int_{\target} (\partial \bar{\Phi})(\partial \Phi)},
\end{split}
\end{equation}
which is just the path integral over a complex scalar field on $\target$.  In this case we choose $[\Phi(x)]^\dag = \bar{\Phi}(x)$ and $[\bar{\Phi}(x)]^\dag = \Phi(x)$ to make the quantum gravity inner product positive-definite.

The generalization of this case to allow general graphs is now straightforward and familiar.  We merely replace the quadratic action in either \eqref{eq:ESTQGA} or \eqref{eq:ESTQGAC} with a more general functional of $\Phi$.  Of course, the Euclidean fields  $\Phi(x)$ remain simultaneously diagonalizable.
Such models have been discussed in the past, see e.g., \cite{Symanzik:1966euc,Nelson:1973mak}.

We can arrive at interesting possibilities in some cases by restricting the allowed set of boundary conditions. Consider a case where we take $\target$ to be Euclidean hyperbolic space $\mathbb{H}_{d}$, and require boundary conditions for the worldlines to end on the asymptotic boundary $S^{d-1} = \partial \mathbb{H}_d$. Take the non-interacting case (without vertices) and with constant potential $U=m^2$. The resulting quantum gravitational amplitudes are the conformally invariant correlators of Euclidean mean field theory\footnote{This depends on a choice of conformal frame (i.e., choice of metric within the conformal class of the boundary $\textbf{S}^{d-1} = \partial \mathbb{H}_d$), related to how we regulate the infinite length of worldlines.}  (or a `generalized free theory') defined on the conformal boundary $\partial \target$ \cite{Maxfield:2017rkn}. We thus arrive at a critical Euclidean statistical theory defined on the boundary of target space. This example is interesting primarily as the free limit of a gravitational theory with AdS asymptotics, for which the boundary Euclidean statistical theory has a local description by the AdS/CFT correspondence: see further discussion in \cref{sec:Disc}.

%~~~~~~~~~~~~~~~~~~~~~~~~~~~~~~~~~~~~~~~~~~~~~~~
\section{Euclidean approach to QFT-like theories}
\label{sec:EQFT}
%~~~~~~~~~~~~~~~~~~~~~~~~~~~~~~~~~~~~~~~~~~~~~~

In \cref{sec:ESTamp}, we discussed how the amplitudes of a one-dimensional theory of Euclidean quantum gravity lead to the correlation functions of a Euclidean QFT. In \cref{sec:ESTBU} we then constructed a Hilbert space from these amplitudes. But the result of our construction was not what one usually calls `the Hilbert space of the QFT'. In this section, we spell out the different choices that lead from the correlation functions \eqref{eq:ESTQGA} to the more familiar notions of QFT Hilbert space. Such constructions have a long history \cite{Schwinger:1958mma}, and are associated in particular with the Osterwalder-Schrader reconstruction theorem \cite{Osterwalder:1973dx,Osterwalder:1974tc} (see e.g., \cite{Glimm:1987ng}). Our main aim here is to understand these ideas in the context of the argument of \cite{Marolf:2020xie} reviewed in \cref{sec:reviewSS}.

 %~~~~~~~~~~~~~~~~~~~~~~~~~~~~~~~~~~~~~~~~~~~~~~~
\subsection{Hilbert space construction}
%~~~~~~~~~~~~~~~~~~~~~~~~~~~~~~~~~~~~~~~~~~~~~~

From the gravitational perspective, since we are taking the same amplitudes as in \cref{sec:EST}, we should ask what other choices were made to construct the Hilbert space, and consider other options. The most obvious is our adjoint operation $\dag$. A simple possible generalization of the choice made in \cref{sec:ESTBU} is to define $\dag$ to act locally on $\target$, so $x^\dag = \sigma(x)$ for some function $\sigma:\target\to\target$. Since the adjoint must square to the identity, so must $\sigma$, so it is required to be an involution: $\sigma(\sigma(x))=x$. This means that $\sigma$ can be represented as a self-adjoint and unitary operator acting on the Hilbert space $L^2(\target)$ of our matter theory. To make the inner product on the one-universe sector conjugate symmetric ($\ASU(\sigma(x),y) = \ASU(\sigma(y),x)^*$), we also require this operator to commute with the matter Hamiltonian $H$, so $\sigma$ should be a $\ZZ_2$ symmetry of $\target$ (preserving the metric and potential). This is of course a constraint on $\target$ as well as $\sigma$, since not every matter theory will admit such a symmetry.

The simplest example to have in mind is the product space $\target = \Sigma\times \RR$ (with a constant potential $U=m^2$), where $\sigma$ acts trivially on $\Sigma$, and reflects the coordinate $t_E$ for the $\RR$ factor: $\sigma(t_E)=-t_E$. We will see that this is the most important example for the application to QFT, since the corresponding amplitudes are the Euclidean correlation functions for the vacuum state of a scalar field of mass $m$ on the spatial manifold $\Sigma$, with $t_E$ interpreted as the Euclidean time. A specific example of this is Bianchi I model with $\target = \mathbb{R}^3$, after Wick rotation $x^0 \to -i\,t_E$.

However, this presents an immediate problem for our inner product on the one-universe Hilbert space. This inner product is computed by the matrix elements of $\sigma \frac{1}{H}$, but this will never be a positive operator, so the inner product will not be positive-definite. The reason is that we can diagonalize $\sigma$ and $H$ simultaneously, so we may choose eigenstates of $H$ with definite parity $\pm 1$ under $\sigma$. But $H$ is a positive operator, so all the eigenstates with negative parity will have negative norm. If all eigenstates have positive parity, it means that $\sigma$ is the identity and we return to the construction of \cref{sec:ESTBU}.

\begin{figure}
\centering
\begin{subfigure}[t]{\textwidth}\centering
	\includegraphics[width=.4\textwidth]{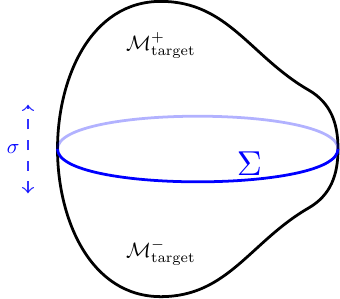}
	\caption{A target space $\target$ with reflection symmetry $\sigma$. The reflection fixes the surface $\Sigma$, which splits the target space into two pieces $\target^\pm$.}
\label{fig:EFTa}
\end{subfigure}
\hfill
\begin{subfigure}[t]{\textwidth}\centering
	\raisebox{30pt}{\scalebox{1.5}{$|x\rangle\sim$}}
	\includegraphics[width=.15\textwidth]{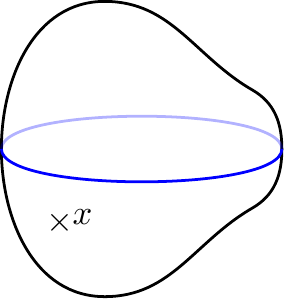}
	\raisebox{30pt}{\scalebox{1.5}{, $|y\rangle\sim$}}
	\includegraphics[width=.15\textwidth]{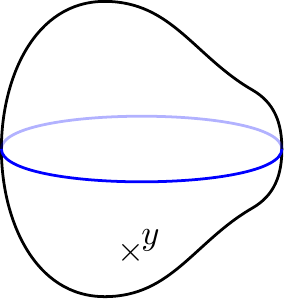}
	\raisebox{30pt}{\scalebox{1.5}{$\longrightarrow\langle y|x\rangle\sim$}}
	\raisebox{-10pt}{\includegraphics[width=.2\textwidth]{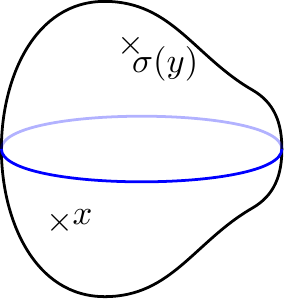}}
	\caption{The single-universe states $|x\rangle$ and $|y\rangle$ are defined by a choice of points $x$ and $y$ (indicated by the crosses $\times$) in the half-space $\target^-$. The inner product $\langle y|x\rangle$ is computed by the worldline path integral, with boundary conditions specifying that worldlines end at  $x$ and $\sigma(y) = y^\dag$. More general states are defined by including multiple insertions in $\target^-$, and most generally by superpositions (both summing over the number of insertions and integrating over their locations with some weighting).}
\label{fig:EFTb}
\end{subfigure}
\caption{\label{fig:EFT}}
\end{figure}

Inspired by constructions of the QFT Hilbert space, we evade this by restricting the set of allowed boundary conditions. Specifically, we restrict attention to the case when the involution $\sigma$ fixes a hypersurface $\Sigma$, and $\target -\Sigma$ has two connected components $\target^-$ and $\target^+=\sigma(\target^-)$. We then define our Hilbert space to be spanned by states $|x_1,\ldots,x_n\rangle$, but now only allowing $x_i\in\target^-$. We illustrate this in \cref{fig:EFT}. In the simple example $\target = \Sigma\times \RR$, the hypersurface $\Sigma$ lies at the moment of time-reflection symmetry $t_E=0$, $\target^+$ and $\target^-$ are the regions $t_E>0$ and $t_E<0$ respectively, and we define states with operator insertions at negative $t_E$ only.

Note that with this restriction on boundary conditions,  our adjoint operation $x^\dag = \sigma(x)$ does not preserve the space $\target^-$ of single-universe boundary conditions to which \eqref{eq:QGIP1} applies. This construction therefore violates an implicit assumption of \cite{Marolf:2020xie}. We will later see some implications for the operators $\hat{x}$.

To see that these choices result in a positive definite inner product and to make contact with QFT constructions, we return to the path integral expression \eqref{eq:ESTQGA} for the amplitudes. We may write our inner product as
\begin{equation}
\label{eq:EQFTIP}
\braket{y_1,\ldots , y_n }{ x_1,\ldots,  x_m} =
	{\cal N} \, \int {\cal D}\Phi \ \Phi(\sigma(y_1)) \cdots \Phi(\sigma(y_n)) \Phi(x_1) \cdots \Phi(x_m)\;
	 e^{-I[\Phi]},
\end{equation}
where $x_1,\ldots, x_m,y_1,\ldots ,y_n$ are points in $\target^-$. To consider general states (linear combinations of states $|x_1,\ldots,x_m\rangle$), we may denote them as functionals of fields as in \eqref{eq:functionalState}, with
\begin{equation}
	\langle G|F\rangle = \mathcal{N} \int \piD{\Phi} \,G^*[\Phi\circ\sigma]F[\Phi]e^{-I[\Phi]},
\end{equation}
but now $F,G$ are functionals depending only on restriction of the field $\Phi$ to $\target^-$. Since  $F[\Phi]$ depends only on the field in $\target^-$ and $G^*[\Phi\circ\sigma]$ only on the field in $\target^+$, we may split the path integral into two pieces, integrating separately over fields $\Phi^\pm$ restricted to the respective regions. These are only identified at the common boundary $\Sigma$, so we have $\left.\Phi^\pm\right|_\Sigma=\Phi_\Sigma$, and the inner product is written as the residual path integral on $\Sigma$:
\begin{gather}
	\langle G|F\rangle =
	\int_\Sigma {\cal D}\Phi_\Sigma  \Psi_G^*[\Phi_\Sigma]	\Psi_F[\Phi_\Sigma] , \label{eq:wavefuntionalIP}  \\
	\text{ where}\quad \Psi_F[\Phi_\Sigma] = \sqrt{\mathcal{N}} \int_{\Phi^-|_\Sigma = \Phi_\Sigma} \piD{\Phi^-} F[\Phi^-] e^{-I_-[\Phi^-]}. \label{eq:wavefunctional}
\end{gather}
The path integral defining $\Psi_F$ is performed over fields $\Phi^-$ on $\target^-$ with the specified values on $\Sigma$, and $I_-$ is the action in \eqref{eq:ESTQGA}, but with the integration restricted to $\target^-$. In this form, the inner product is manifestly positive semidefinite, since the norm of the state $F$ is given by integrating the positive functional $|\Psi_F(\Phi_\Sigma)|^2$ with respect to a positive measure.\footnote{Interestingly, it is more complicated to show positive-definiteness working directly within the worldline formalism. This is related to the comment in \cite{Marolf:2020xie} that it is unclear what conditions on the quantum gravity path integral are required for positive-definiteness of the full quantum gravity inner product.  We include a worldline argument for positive-definiteness of the inner product in \cref{app:WPD} for the Gaussian case.}

Indeed, we can interpret \eqref{eq:wavefunctional} as the path integral computation of a Schr\"odinger-picture wavefunctional on the surface $\Sigma$, and \eqref{eq:wavefuntionalIP} as the inner product of two such wavefunctionals. In the example $\target = \Sigma\times \RR$, the no-boundary state $|\HH\rangle$ constructs the vacuum of the QFT at time $t_E=0$, and nontrivial boundary conditions for worldlines produce excited states by inserting (Euclidean time-ordered) operators in the lower half-space $t_E<0$. Our choices are essentially equivalent to the Osterwalder-Schrader construction of the QFT Hilbert space on the spatial slice $\Sigma$ from Euclidean correlation functions.

 %~~~~~~~~~~~~~~~~~~~~~~~~~~~~~~~~~~~~~~~~~~~~~~~
\subsection{Null states}
%~~~~~~~~~~~~~~~~~~~~~~~~~~~~~~~~~~~~~~~~~~~~~~

Unlike in \cref{sec:ESTBU}, the inner product defined above admits nontrivial null states. As observed in \cite{Anous:2020lka}, we may think of these as arising from the QFT field equations, which we may here write as $H\Phi(x)=0$ where $H$ is the matter Hamiltonian for our one-dimensional quantum gravity theory. Importantly, this holds only at separated points: when field insertions collide this may be violated by contact terms. Indeed, these contact terms explain why the field equations did \emph{not} give rise to null states in \cref{sec:ESTBU}. However, with our restricted boundary conditions $x,y\in\target^-$, operators $\Phi(x)$ and $\Phi(\sigma(y))$ can never collide, so inner products do not produce contact terms.

More explicitly, let us focus on the one-universe Hilbert space, and consider states $|f\rangle = \int dx\, f(x) |x\rangle$, where $f$ has compact support contained in $\target^-$. In particular, we may choose a function $f = H h$, where the support of $h$ is contained in $\target^-$.
Now the inner product with another state $|g\rangle= \int dx\, g(x) |x\rangle$ is given by
\begin{equation}\label{eq:nullEQFT}
	\langle g|f\rangle = \langle g\circ\sigma |\frac{1}{H}|f\rangle_\mathrm{matter} = \langle g\circ\sigma |h\rangle_\mathrm{matter} =0,
\end{equation}
where $|f\rangle_\mathrm{matter}$ is the state in the matter Hilbert space with wavefunction $f$ (with an $L^2$ inner product on $\target$). The final inner product vanishes because the support of $h$ and $g\circ\sigma$ are disjoint, lying within $\target^-$ and $\target^+$ respectively. Hence, $|f\rangle$ has vanishing inner product with any state, so it must be null: $|f\rangle=0$. To connect this with the field equations $H\Phi(x) =0$, the state is associated with the functional $\int dx\, H h(x) \Phi(x)=\int dx\, h(x) H\Phi(x)$ (where we may `integrate by parts' because $H$ is a symmetric operator on $L^2(\target)$).

While at first the states of the one-universe Hilbert space would appear to be determined by functions $f$ on $\target^-$, such null states mean that the independent states in $\hbu$ are in fact determined by far less data: a function only on $\Sigma$ (see appendix \ref{app:WPD} for a non-redundant characterization of states in terms of a function $\Sigma\to \CC$). Thus, the QFT Hilbert space constructed in this section is in a sense much smaller than that discussed in \cref{sec:EST}, with one-universe states determined by a function on the submanifold $\Sigma$ of $\target$.

%~~~~~~~~~~~~~~~~~~~~~~~~~~~~~~~~~~~~~~~~~~~~~~~
\subsection{Operator algebra}
%~~~~~~~~~~~~~~~~~~~~~~~~~~~~~~~~~~~~~~~~~~~~~~

Finally, we briefly discuss the construction of operators in this formalism. From the general discussion of \cref{sec:reviewSS}, one might expect to have boundary-inserting operators $\hat{x}$ acting on the baby universe Hilbert space. But in fact, these operators are not well-defined, since they do not preserve the space of null states. For example, consider acting on the null state $|f\rangle$ with $f=Hh$ in \eqref{eq:nullEQFT} with $\hat{x}$, inserting a boundary at a point $x\in \target^-$, and take the overlap with the no-boundary state:
\begin{equation}
	\langle \HH | \hat{x} |f\rangle  = \langle x|\frac{1}{H}|f\rangle_\mathrm{matter}= \langle x|h\rangle_\mathrm{matter} = h(x).
\end{equation}
This will be nonzero for some choice of $h$, so $\hat{x}|f\rangle$ is not a null state. As a result, the boundary-inserting operator $\hat{x}$ does not give a well-defined operator on the baby universe Hilbert space, where we have performed a quotient by null states.

Such a result was possible only because our Hilbert space does not obey the axioms of \cite{Marolf:2020xie}. The particular failure is the invariance of the set of allowed boundary conditions under the CPT operation $\dag$, which means that the argument of \eqref{eq:null} is inapplicable.

This is in fact perfectly in line with expectations from QFT.\footnote{This is associated with the well-known fact that while the Osterwalder-Schrader reconstruction directly reconstructs the states of QFT, it does not provide a similarly direct construction of the operator algebra.} We would expect the operators $\hat{x}$ to be associated with the quantum fields $\hat{\Phi}(x)$, but these do not give a well-defined operator algebra on a Euclidean space. For example, if $\target = \Sigma\times \RR$ and we use the usual quantization with respect to Euclidean time $t_E$, products of field operators are sensible only when they appear in Euclidean time order.\footnote{More precisely, the domain of $\hat{\Phi}(x)$ and the image of $\hat{\Phi}(y)$ are disjoint if $x$ is in the Euclidean future of $y$ (except in the case when $\target$ is one-dimensional).}\footnote{There are some boundary-inserting operators $\hat{x}$ that are well-defined (though unbounded) on the baby universe Hilbert space, namely when $x$ lies on the $\sigma$-invariant slice $\Sigma$. Since this set is invariant under our CPT operation, the general argument of \cite{Marolf:2020xie} applies. And indeed, these operators are self-adjoint and mutually commuting, hence simultaneously diagonalizable (with eigenstates corresponding to delta-function wavefunctionals in \eqref{eq:wavefuntionalIP}).}

%~~~~~~~~~~~~~~~~~~~~~~~~~~~~~~~~~~~~~~~~~~~~~~~
\subsection{Generalizations}
%~~~~~~~~~~~~~~~~~~~~~~~~~~~~~~~~~~~~~~~~~~~~~~
\label{sec:EQFTgeneralizations}

This above discussion is readily generalized to any context where quantum field theory is well-defined and has the required ${\mathbb Z}_2$ reflection symmetry.  For example, we may discuss complex scalar fields by using oriented worldlines, or generalize the amplitudes to general Feynman graphs in the manner of interacting quantum field theory.

Note, however, that gravitational models will not typically admit a time-reflection symmetry (for example, the Kantowski-Sachs model introduced later in \eqref{eq:HKS}). Indeed, a $\ZZ_2$ symmetry reversing time would usually be a relation between the physics of large universes and of small universes, so there are few semiclassical gravitational models to which this formalism could apply.  However, it might be interesting to consider whether --- at least in some cases --- string-inspired models with some suitable notion of T-duality (see e.g.\ \cite{Polchinski:1998rr}) could provide the required symmetry.

%~~~~~~~~~~~~~~~~~~~~~~~~~~~~~~~~~~~~~~~~~~~~~~~
\section{Group Averaged Theories}
\label{sec:GAT}
%~~~~~~~~~~~~~~~~~~~~~~~~~~~~~~~~~~~~~~~~~~~~~~

We  now turn  to the case where the worldline is taken to have Lorentz signature. We will begin by discussing a framework that may be unfamiliar to many practitioners of QFT or string theory. It is however inspired by a popular approach to studying single-universe quantum gravity models by treating them as constrained systems \cite{Landsman:1993xe,Marolf:1994wh,Ashtekar:1995zh,Marolf:1996gb,Reisenberger:1996pu,Hartle:1997dc,Marolf:2000iq,Shvedov:2001ai}  and by treatments \cite{Higuchi:1991tm} of linearization-instabilities in quantum gravity.

The single-universe amplitudes $\ASU(x,y)$ that enter the quantum gravity amplitudes \eqref{eq:QGA}  take the form
\begin{equation}
\ASU(x,y) = \int_{\cal D} dT \langle x | e^{-iHT} | y \rangle_\mathrm{matter}
\end{equation}
for some choice of integration domain ${\cal D}\subset {\mathbb R}$.  The three options respecting worldline locality are the positive reals ${\mathbb R}^+$, the negative reals ${\mathbb R}^-$, or the full real line $\mathbb{R}$.
We here discuss the latter choice  (${\cal D} = {\mathbb R}$), which fits into the so-called `group averaging' paradigm discussed in \cite{Higuchi:1991tm,Landsman:1993xe,Marolf:1994wh,Ashtekar:1995zh,Marolf:1996gb,Reisenberger:1996pu,Hartle:1997dc,Marolf:2000iq,Shvedov:2001ai}
 (sometimes under other names).  With this choice, we may write
\begin{equation}
\label{eq:ASUGA}
\ASU(x,y) = \mel{x}{ \delta(H) }{ y}_\mathrm{matter} \,,
\end{equation}
which shows that choosing ${\cal D} = {\mathbb R}$ imposes the constraint $H=0$ in a strong sense.\footnote{This raises the issue that is sometimes called the problem of time in quantum gravity, which is resolved by realizing that the true dynamics of such systems are encoded in relational observables.  See e.g. \cite{DeWitt:1962cg,DeWitt:1967yk,Rovelli:1990jm,Rovelli:1989jn,Rovelli:1990ph,Smolin:1993ka,Marolf:1994wh} for classic discussions.}

As an example, consider  the mini-superspace truncation of the Bianchi I model discussed in  \cref{sec:general}, where the
Hamiltonian \eqref{eq:rescaledB1} (in rescaled lapse variable) corresponds to that of a free particle motion in $\mathbb{R}^{2,1}$.  The single universe amplitudes compute the matrix elements of the on-shell  constraint $\delta(p^2)$.  Any general linear combination
$\int_{\mathbb{R}^{2,1}} f(y)  \ket{y}$ of the allowed boundary conditions is effectively projected into the space of solution of the quantum Hamiltonian constraint $H\ket{\psi} =0$.

From \eqref{eq:ASUGA} it is manifest that our single-universe amplitudes are the matrix elements of a positive operator on the matter Hilbert space.  As a result, we can define a positive-definite quantum gravity inner product by taking $\CPT$ to act trivially on $\target$; i.e., $x^\dag = x$.    In contrast, integrating $T$ only over a half-line would generate complex single-universe amplitudes from which the construction of a good Hilbert space is more complicated (see \cref{sec:LQFT}).

Since \eqref{eq:ASUGA} projects onto solutions of the constraint $H|\psi\rangle =0$, we may obtain a more explicit description of our theory by working directly with such solutions. In particular,  for the Bianchi I model since
\begin{equation}
\delta(p^2) = \frac{1}{|p_0|} \,\delta\left(p_0 - \sqrt{p_1^2+p_2^2}\right) + \frac{1}{|p_0|}\, \delta\left(p_0 + \sqrt{p_1^2+p_2^2}\right),
\end{equation}
it suffices to use the plane wave solutions
\begin{equation}
\braket{ x}{p_1,p_2;\,\eta } = e^{- i\, \eta\, x^0\, \sqrt{p_1^2 + p_2^2}} \ e^{i(x^1p_1+x^2p_2)}
\end{equation}
for $\eta =\pm$ and to replace \eqref{eq:ASUGA} by the `projected' amplitudes
\begin{equation}
\label{eq:projASUGA}
\tASU(p',p) = \frac{1}{|p_0|} \, \delta(p_1-p_1')\,\delta(p_2-p_2')\, \delta_{\eta \eta'},
\end{equation}
where on the left-hand side $p = (p_1,p_2,\eta)$ and $p'=(p_1',p_2',\eta')$.
In particular, we emphasize that the group-averaging approach keeps both positive- and negative-frequency solutions to the constraint and treats both on an equal footing.

The full quantum gravity Hilbert space can now be described succinctly by using the observation that \eqref{eq:QGA} is just the result of performing a Gaussian integral over the space of plane wave solutions $p$ with covariance given by the right-hand side of \eqref{eq:projASUGA}.  Consistent with the general argument from \cite{Marolf:2020xie}, the allowed boundary conditions $p$ then define a set of simultaneously-diagonalizable operators on this Hilbert space.

%~~~~~~~~~~~~~~~~~~~~~~~~~~~~~~~~~~~~~~~~~~~~~~~
\subsection{Generalizations}
%~~~~~~~~~~~~~~~~~~~~~~~~~~~~~~~~~~~~~~~~~~~~~~
\label{sec:GATgeneralizations}

The above construction can be used in great generality.  As it clear from \eqref{eq:ASUGA}, it requires only a matter Hamiltonian $H$ with continuous spectrum that includes zero and whose matrix elements are symmetric (so that $\ASU(x,y) = \ASU(y,x)$). The latter requirement is tantamount to requiring the matter to have a time-reversal symmetry, and is a consequence of our summing over unoriented spacetimes.  This condition would be dropped if we instead summed over oriented spacetimes.

In particular, the above conditions on $H$ allow the case of sigma-models with general (geodesically complete) Lorentzian target space, or in fact any signature with additional `time' directions (or indeed none) so long as the sign of $H$ remains indefinite.    We are similarly free to add a potential to $H$ that preserves continuity of the spectrum and the inclusion of the eigenvalue zero.  Using a rescaled notion of lapse  as described in \cref{sec:minisuper}, this allows one to treat rather general minisuperspace models \cite{Marolf:1994wh,Marolf:1994ss}.

Of perhaps greater interest is the generalization to  allow graphs instead of summing only over strict one-dimensional manifolds.  This might be expected to reproduce more the features of higher-dimensional gravity that one might expect from the diagrams of  \cref{fig:wormholes}.  Since this is not the focus of the current paper, we will content ourselves by noting that some such generalizations are straightforward.  For example, in writing down the final quantum gravity Hilbert space one can readily replace the Gaussian integral over solutions to the constraint $H|\psi\rangle=0$ with some non-Gaussian integral over the same space of constraints.  Similarly, one can allow the constraint to depend on the `coupling constants' $g$ that control any non-Gaussianities, so that the space of random variables at each $g$ is defined by solving a new constraint $H_g|\psi \rangle =0$.  In each case, the allowed boundary conditions continue to define simultaneously diagonalizable sets of operators.  We leave for future investigation whether the diagrammatics of such theories matches expectations from the quantum gravity path integral, though we will return to comment further on the physics of quantum gravity constraints in \cref{sec:Disc}.

%~~~~~~~~~~~~~~~~~~~~~~~~~~~~~~~~~~~~~~~~~~~~~~~
\section{Lorentzian approach to QFT-like theories}
\label{sec:LQFT}
%~~~~~~~~~~~~~~~~~~~~~~~~~~~~~~~~~~~~~~~~~~~~~~

The construction of \cref{sec:GAT} led us to a Hilbert space for (amongst other things) worldline gravity with Lorentz-signature target space. For example, the matter Hamiltonian of the Bianchi I model is given by the wave operator on 3-dimensional Minkowski spacetime. However, we did not arrive at the quantum field theory Hilbert space (for a free field on $\RR^{2,1}$ in the Bianchi I case, for example) as one might have expected form the worldline formalism of QFT \cite{Strassler:1992zr,Schubert:2001he}. In this section we examine the choices one might take (different from \cref{sec:GAT}) to pursue the analogy with QFT.

We first note that the amplitudes defined from our GAT are not the usual correlation functions of QFT. The one-universe amplitudes \eqref{eq:ASUGA} are the matrix elements of $\delta(H)$, while the usual two-point functions of free fields are Green's functions of the matter Hamiltonian (inverses of $H$). This is a result of integrating the lapse $T$ over the entire real line, $\mathcal{D}=\RR$. We can consider instead the choice of integrating only over a half-line, say  $\mathcal{D}=\RR^+$:

\begin{equation}
\label{eq:ASUQFT}
\ASU(x,y) = \int_{{\mathbb R}^+} dT \mel{x}{e^{-iHT} } {y}= \mel{x}{  \frac{1}{i(H-i\epsilon)} }{ y},
\end{equation}
Here the integral gives a well-defined a distribution (associated with the Fourier transform of a step function) which we have written in terms of an $i\epsilon$ prescription. This is in fact gives a standard description of $iG_F$ where $G_F$ is the Feynman Green's function (see e.g., \cite{dewitt1965dynamical}).  In particular, the path integral is  well-defined for Hamiltonians with a continuous spectrum, which need not be bounded, and computes matrix elements between states with an appropriately smooth and rapidly-decaying energy representation.   The result can also be written in the form
\begin{equation}
\label{eq:pole}
	\frac{1}{i(H-i\epsilon)} = \mathcal{P}\frac{1}{i H} + \pi\, \delta(H),
\end{equation}
where $\mathcal{P}$ denotes the principal value distribution.

Due to the imaginary part of \eqref{eq:pole}, using \eqref{eq:QGIP1} as written, and taking $\CPT$ acting trivially on ${\cal M}_{\text{target}}$, one no longer defines a real and positive quantum gravity inner product.  Yet QFTs do have a positive inner product as well as a notion of  $ \CPT$-conjugation (e.g., for real scalar fields $\Phi(x)^\dagger = \Phi(x)$).  Furthermore, in the free case that is relevant when we exclude non-trivial graphs, it has precisely the structure described by \eqref{eq:QGA}.

The explanation is that in QFT the amplitude \eqref{eq:ASUQFT} does not define an inner product between states $\Phi(x)|\Omega\rangle$ and $\Phi(y)|\Omega\rangle$ for some `vacuum' $|\Omega\rangle$. That inner product would be given by the expectation value of $\Phi(y)\Phi(x)$, with operators ordered as written (a Wightman function). But \eqref{eq:ASUQFT} would instead be interpreted as a time-ordered correlation function: the expectation value of $\mathcal{T} \left\{\Phi(x)\Phi(y)\right\}$, where the ordering of operators depends on their order in Lorentzian time. Choosing the lapse to run over the negative reals gives us the anti-time-ordered correlation function (the complex conjugate of the time-ordered correlator). In this QFT language, our choice of integrating over the entire real line in \cref{sec:GAT} gives the sum of these, which is the expectation value of the anti-commutator $\Phi(x)\Phi(y)+\Phi(y)\Phi(x)$.

To recover the QFT (Wightman) correlation functions, we must therefore apply \eqref{eq:ASUQFT} only when $y$ lies to the future of $x$, and instead use its conjugate (integrate over $T\in\RR^-$) if $x$ lies to the past of $y$. But this condition is not symmetric in $x$ and $y$: the unordered list of boundary conditions $x$ is not sufficient information to tell us to construct the correlation function, as in the formalism of \cref{sec:reviewSS} (in particular \ref{eq:BUperm}).

When the target space $\target$ is Minkowski space, we recover the familiar correlation functions of free QFT in the vacuum state, and may construct from them the associated Hilbert space \cite{Wightman:1956zz}. However, this construction of a Hilbert space does not work for general $\target$, since it is not guaranteed that \eqref{eq:ASUQFT} will be consistent with a time-ordered correlation function in any state.\footnote{We thank E.~Witten for discussions bringing this point to light.} The problem is that the correlation function $\langle \mathcal{O}^\dag \mathcal{O}\rangle$ may not be positive (or even real) for general smearings $\mathcal{O} = \int f(x) \Phi(x) dx$ of $\Phi$, even once we interpret \eqref{eq:ASUQFT} as a time-ordered correlator. To construct a Hilbert space requires some condition on $\target$ to ensure positivity in this sense, for example that $\target$ admits a time-reversal symmetry. In such a case, we obtain not only the Hilbert space of free QFT, but also a particular Gaussian state $|\Omega\rangle$ of the theory.

With these choices, the Hilbert space becomes the usual bosonic Fock space associated with a real scalar fields  on $\target$.  We then find operators $\hat \phi(x)$ associated with each boundary condition $x \in \BC = \mathbb{R}^{2,1}$, but their algebra is famously non-abelian and the operators cannot be simultaneously diagonalized.  We see that this is a result of deviating from the structure assumed in the arguments of \cite{Marolf:2020xie}.  Of course, one may nevertheless choose to focus on some abelian subalgebra, perhaps the one defined by choosing\footnote{Note that restricting to a constant value of $x^0$ means that one can work entirely with \eqref{eq:ASUQFT} without requiring the complex conjugate.  As a result, with this restriction the formalism fits within the framework of \cite{Marolf:2020xie}, and the resulting abelian algebra is consistent with the general argument of \cite{Marolf:2020xie}.} $x^0=0$.  This seems to be the approach taken in \cite{Coleman:1988cy,Giddings:1988cx,Polchinski:1994zs}, though the more general `3rd quantization' described in \cite{Giddings:1988wv} allows the algebra of operators on $\hbu$ associated with boundary quantities to be non-abelian.

For an alternative approach that remains somewhat closer to the axiomatic framework of \cite{Marolf:2020xie}, one may assign an additional parameter to each boundary condition that determines their relative ordering. We must then define an adjoint operation $\dag$ that reverses that ordering, and additionally place a constraint on boundary conditions so that `bra' boundaries are always ordered after `ket' boundary conditions. One can think of this as formally assigning each operator with an `$i \epsilon$' deformation in imaginary time, with correlation functions always in Euclidean time order. The adjoint takes $\epsilon\to-\epsilon$, and we restrict boundary conditions to negative $\epsilon$. The result is much like the Euclidean construction in \cref{sec:EQFT}, and  in particular the structure of null states and the operator algebra is similar. Equivalently, we may think of the target space as having many `time-folds' in a Schwinger-Keldysh type contour, and our boundary conditions label which contour an operator insertion lies on (cf., \cite{Haehl:2017qfl}). Correlation functions are then always contour-ordered, the adjoint reverses the order of contours, and we restrict our states to be defined by insertions on the appropriate half of the contours.

%~~~~~~~~~~~~~~~~~~~~~~~~~~~~~~~~~~~~~~~~~~~~~~~
\subsection{Generalizations}
%~~~~~~~~~~~~~~~~~~~~~~~~~~~~~~~~~~~~~~~~~~~~~~
\label{sec:LQFTgeneralizations}

%Note first that  interpreting our amplitudes \eqref{eq:ASUQFT} as time-ordered correlation functions uniquely constructs not only the Hilbert space, but also a particular state $|\Omega\rangle$ in the theory.  It is natural to expect that this $|\Omega\rangle$ is the Sorkin-Johnston state of \cite{Sorkin:2011pn,Johnston:2009fr,Afshordi:2012jf}, as that state is also defined covariantly from the QFT equation of motion.  However, we save investigation of this potential connection for future work.

We expect that this approach can be generalized significantly. However, to maintain contact with a worldline quantum gravity formalism one would like to continue to use \eqref{eq:ASUQFT} (or its conjugate) to define the single-universe amplitudes.  An important question is then to understand the class of models for which the resulting inner product is positive definite. When this is the case, we expect that our procedure defines the usual Hilbert space for the associated free quantum field theory.\footnote{The idea is that, so long as the metric and any potential are smooth, in the far UV our construction will agree with the case of Minkowski-space QFT.  This is well-known to give the correct UV behavior for any QFT.  And, in the absence of IR divergences, this condition determines a unique space of QFT states; see e.g  \cite{Wald:1995yp}.  So if our procedure defines a Hilbert space, it must be the usual one of QFT.}

Nevertheless, at least a few minimal properties would seem to be required for success.
The first is that there be some notion of target space $\target$, where in particular $\target$ is a time-orientable Lorentz-signature manifold. While the definition of the  amplitude in \eqref{eq:ASUQFT} makes sense in any signature, only time-orientable Lorentzian spacetimes have a partial order imposed by the causal structure. This partial order is required to interpret the amplitudes as expectation values of time-ordered products, and hence would appear to be essential to construct a QFT-like Hilbert space. In addition, in Minkowski space the discussion is simplified by the fact that $H$ is an essentially self-adjoint operator on $L^2(\target)$.  When this is not the case, the role of boundary conditions defined by the details of the matter path integral will be more important, and we expect it to be necessary to choose boundary conditions that make $H$ self-adjoint.  Essential self-adjointness is to be expected when $\target$ is both globally hyperbolic and geodesically complete, though more generally it should be expected to fail.

In addition to such mathematical questions, additional physical considerations may be relevant as well.  For example, as pointed out long ago in \cite{Marolf:1994wh}, the physics appropriate for a QFT may not always agree with the physics appropriate for a theory of quantum gravity.  To illustrate this issue, let us follow \cite{Marolf:1994wh} and consider a gravitational model in which universes start from a big bang, expand to some maximum size, and then collapse to a big crunch.  A simple example is given by the so-called Kantowski-Sachs model of anisotropic vacuum gravity on ${\bf S}^1 \times S^2 \times \RR$, which in the end \cite{Ashtekar:1993wb} differs from \eqref{eq:rescaledB1} only by using symmetry to set $x_2=0$ and adding an external potential:
\begin{equation}
\label{eq:HKS}
H = -p_0^2 + p_1^2 - 48 \,e^{2 \, (2 \, x^0- x^1)}.
\end{equation}
The overall classical dynamics is clear from the fact that any future-directed timelike curve on ${\cal M}_{\text{target}} = \mathbb{R}^{1,1}$ has increasing $2\,x^0 - x^1$.  So classical solutions cannot be completely described by such future-directed timelike curves, as in the far future this would require the spacelike condition $p_1^2 - p_0^2 > 0$.  Instead, while one might begin with such a curve (say, emerging from a big bang at $x^0 = -\infty$), the dynamics forces the trajectory to become spacelike at some point, and in fact to turn around so that $x^0$ then begins to decrease.  Eventually, the trajectory becomes time-like again, but it is now past-directed on ${\cal M}_{\text{target}} = \mathbb{R}^{1,1}$ and eventually results in a big crunch at $x^0 = -\infty$.

One might thus expect the quantum version of this model to display similar behavior.  But the quantum field theory associated with \eqref{eq:HKS} is very different.  The scalar experiences an external potential which becomes unboundedly negative at large positive $x^0$.  This results in enormous particle creation, which in a gravitational interpretation would imply creation of a large number of universes at large $x^0$.  Thus, rather than large $x^0$ (scale factor) being forbidden as in the classical theory, large scale factor seems to be dynamically preferred in the QFT-like quantum treatment.

%~~~~~~~~~~~~~~~~~~~~~~~~~~~~~~~~~~~~~~~~~~~~~~~
\section{Discussion}
\label{sec:Disc}
%~~~~~~~~~~~~~~~~~~~~~~~~~~~~~~~~~~~~~~~~~~~~~~

In the above, we have enumerated various choices that could be made in order to build a theory of one-dimensional quantum gravity from worldline path integrals.  We concentrated on situations where splitting and joining of universes is forbidden, so that the final quantum gravity inner product could be represented by a Gaussian integral.  However, generalizations to include interactions between worldlines were briefly discussed in sections \ref{sec:ESTgeneralizations}, \ref{sec:EQFTgeneralizations}, \ref{sec:GATgeneralizations}, and \ref{sec:LQFTgeneralizations}. At least in perturbation theory the addition of such interactions cannot change our qualitative results.  

The various choices that we examined are summarized in \cref{sec:intro} in  \cref{tab:choices}, along with the particular options that lead to  Euclidean Statistical Theories (ESTs), Group Averaged Theories (GATs), and to QFT-like theories.  Both ESTs and GATs fall within the framework described in \cite{Marolf:2020xie}, and in particular define the quantum gravity inner product from the path integral with no restriction on boundary conditions.  Since the GATs are Lorentzian and the ESTs are Euclidean, this demonstrates that the framework accommodates path integrals defined by spacetimes of either signature.\footnote{See also the recent Lorentz-signature discussion of baby universes and superselection in \cite{Marolf:2020rpm}.}  In contrast, in either signature, the QFT-like approaches relate the final inner product to path integral amplitudes only with some restriction on the possible boundary conditions which is not invariant under the CPT operation used to define the inner product. Since this restriction violates the framework described in \cite{Marolf:2020xie}, for such cases path integral boundary conditions need not define simultaneously-diagonalizable operators and hence also do not lead to superselection sectors.

Of course, given a non-abelian algebra of operators, it is possible to select an abelian sub-algebra.  As a result, even using a QFT-like approach, one could attempt to claim that natural boundary objects correspond to such a sub-algebra of the possible baby universe operators;  one may thus consider them to be superselected.  This appears to be the approach that was taken in \cite{Coleman:1988cy,Giddings:1988cx}.  However, as emphasized in \cite{Giddings:1988wv}, such approaches involve assumptions about the nature of the final results --  in particular, a certain form of locality was assumed in \cite{Coleman:1988cy,Giddings:1988cx}.  Reasoning of this kind thus lead \cite{Giddings:1988wv} to question the supposed locality, and to suggest that strict superselection may not hold.  This stands in sharp contrast with the treatment of \cite{Marolf:2020xie} reviewed in \cref{sec:reviewSS}, which argued that properties of the quantum gravity path integral require \emph{all} boundary quantities to be simultaneously diagonalizable in $\hbu$.

By describing the above choices, we hope to reduce confusion in the literature.  For example, the comments of \cite{McNamara:2020uza} concerning 2d theories of gravity arising from Wilson loops in Yang-Mills theories appear to be couched in the EST framework; it is unclear to us whether analogous comments hold in GAT-like constructions.

\paragraph{Implications for quantum gravity:}
In comparing the options described here, the idea that general path integral boundary conditions define the quantum gravity inner product appears natural from many perspectives and is related to historic discussions of quantum gravity and constrained systems
\cite{Halliwell:1989dy,Halliwell:1990qr,Higuchi:1991tm,Landsman:1993xe,Marolf:1994wh,Ashtekar:1995zh,Marolf:1996gb,Reisenberger:1996pu,Hartle:1997dc,Marolf:2000iq,Shvedov:2001ai}; see also \cite{Jafferis:2017tiu}. This was the point of view taken in \cite{Marolf:2020xie}, and was used as the rationale there to justify the assumptions made in that work.  However, we also noted that one may adapt arguments from \cite{Marolf:1994wh} to show that -- at least for certain models of gravitational physics -- applying a QFT-like approach to quantum gravity leads to physics very different from the classical theory, even in an apparently semiclassical domain.  In particular, for any cosmological model in which the universe always reaches a maximum size and then recollapses, QFT-like approaches lead to divergent `pair production' of universes with very large size.  Thus, while the classical physics forbids universes of arbitrarily large size, the physics of QFT-like quantum gravity would be dominated by arbitrarily large universes.  In contrast, the physics of GAT quantum gravity for those models again forbids large universes and is thus consistent with the classical results.  We take this as a further argument against the use of QFT-like constructions in quantum gravity.

\paragraph{Other discrete choices:}
While we have explored several important discrete choices in the construction of candidate quantum gravity theories, it is important to emphasize that there are many other places at which further choices could be introduced and  whose consequences remain to be explored.  For example, as described in \cref{sec:reviewSS} our quantum gravity amplitudes were taken to be completely symmetric.  This naturally reminds one of bosonic quantum field theory, and thus leads to the question of whether other forms of multi-universe statistics might be allowed and what consequences they might entail. One could also as noted include worldline supersymmetry. Similarly, for simplicity we considered only the case of unoriented spacetimes (worldline), while summing instead of oriented worldlines should lead to slight modifications in parallel with the two-dimensional discussion in \cite{Stanford:2017thb}.

Indeed, many additional choices naturally arise when one generalizes the theory to allow splitting and joining of universes by summing over graphs.  For example, in that context one may elect to treat `internal' lines differently than `external' lines.  One may also allow the matter Hamiltonian $H$ to have some explicit dependence on the `coupling constants' $g$ associated with the graph vertices, or perhaps to further generalize the structures discussed above.

\paragraph{Interpretation of low dimensional topological models:}
However, the most intriguing questions that stem from this work revolve around the relationship between the models discussed here and the topological model of \cite{Marolf:2020xie}, the treatments of JT gravity in \cite{Saad:2018bqo,Saad:2019lba,Blommaert:2019wfy,Saad:2019pqd,Penington:2019kki,Marolf:2020xie,Blommaert:2020seb,Bousso:2020kmy,Stanford:2020wkf}, and higher dimensional gravity more generally.  For example, having noted that our EST and GAT constructions both fall within the general framework described in \cite{Marolf:2020xie}, one might wonder which best corresponds to the way in which the topological model of \cite{Marolf:2020xie} was explicitly solved in that work.   However, one should recall that the main differences between ESTs and GATs involved the treatment of the constraint $H$ and the associated integration domain for the proper time $T$, and that neither $H$ nor $T$ appears at all in a topological model like that considered in \cite{Marolf:2020xie}.  As a result, it is far from clear whether such models can be meaningfully associated with either construction.

Now, in considering either the topological model of \cite{Marolf:2020xie} or the treatments of JT gravity in \cite{Saad:2018bqo,Saad:2019lba,Blommaert:2019wfy,Saad:2019pqd,Penington:2019kki,Marolf:2020xie,Blommaert:2020seb,Bousso:2020kmy,Stanford:2020wkf}, one might note that all of these works focus on Euclidean path integrals and thus be tempted to associate them with the construction of ESTs in \cref{sec:EST} in contrast to the Lorentzian construction of GATs in \cref{sec:GAT}.  However, in \cref{sec:EST} and \cref{sec:GAT}, the terms Lorentzian and Euclidean were used to describe the natural contours of integration that {\it define} the desired path integral, while in many other contexts one uses the term to describe the sorts of spacetimes that one uses to evaluate the path integral.  This distinction is illustrated by standard non-relativistic quantum mechanics, which one may choose to think is fundamentally defined by a real-time ('Lorentz signature') path integral, but which is usefully evaluated using Euclidean methods in contexts that involve quantum tunneling under a classical barrier.  In essence, the point is that one may typically deform the contour of integration into the complex plane to rewrite an originally-Lorentzian path integral in a Euclidean form, or to
rewrite an originally-Euclidean path integral in a Lorentzian form.\footnote{One may also note that, even in standard quantum mechanics, one is free to study `Euclidean' quantities like $e^{-HT}$ in the `Lorentzian' theory. }
And it was shown in \cite{Marolf:1996gb} that this could be done for the GAT path integral by taking the contour to rotate in different directions depending on the particular matter boundary conditions chosen.

As a result, the mere use of Euclidean techniques in the above references is not sufficient to conclude that their treatment is more closely related to our ESTs than to our GATs.  Indeed, we note that at least one element of the treatment in \cite{Saad:2018bqo,Saad:2019lba,Blommaert:2019wfy,Saad:2019pqd,Penington:2019kki,Marolf:2020xie,Blommaert:2020seb,Bousso:2020kmy,Stanford:2020wkf} appears to resemble the GAT construction.  Namely, motivated by the fact that the dilaton $\phi$ appears linearly in the JT gravity action, the above references take functional integration over the dilaton to exactly enforce the metric equation of motion $R=-2$.  Thus they choose an integration contour for $\phi$ much like the GAT contour for the proper time $T$ (which resulted in exactly enforcing the equation of motion $H=0$ by producing a $\delta(H)$).  However, we leave for future work any more detailed analysis of this JT path integral prescription and possible connection with GATs (or with generalizations thereof).

\paragraph{Minisuperspace models of quantum gravity:}
Indeed, an important question is the extent to which physics of higher-dimensional quantum gravity is in fact similar to {\it any} of the models described here.  We consider this to be an open question, with much to be investigated.  In particular, in our work above, we took the matter Hamiltonian $H$ to be a self-adjoint operator.  While this seems natural from the perspective of familiar matter quantum mechanics, we believe it to be rather less obvious from the perspective of higher-dimensional quantum gravity.  To this end we remind the reader that when one defines a one-dimensional quantum gravity model by Kaluza-Klein reduction of a higher-dimensional theory, most of the features and complications of the gravitational theory then become part of the Kaluza-Klein matter sector, with only some overall notion of proper time left to be treated as one-dimensional gravity.

Let us thus consider again the diagrams of  \cref{fig:wormholes} associated with the splitting and joining of universes.  From the higher-dimensional perspective, these are smooth Euclidean geometries.  And in general, one expects smooth Euclidean solutions to be associated with tunneling amplitudes between e.g., classically-allowed Lorentzian configurations.

We can discuss this in more detail in a simple (and famous) minisuperspace model associated with the Hartle-Hawking wavefunction of the universe.  To this end, consider spatially compact homogeneous isotropic universes with topology ${\bf S}^3 \times {\mathbb R}$ in the presence of a {\it positive} cosmological constant, but with no explicit matter.  Such cosmologies are described by a minisuperspace model having a single degree of freedom $a$ (`the scale factor'), which upon Kaluza-Klein reduction becomes our matter sector.  The scale factor takes values in ${\mathbb R}^+$, and the corresponding `matter' Hamiltonian generating evolution in proper time is then $H = p_a^2 + V(a)$, with $V(a) = 1- \frac{4}{3}\, a^2$. Note that $H$ is {\it not} essentially self-adjoint, as classical solutions associated with energies greater than unity reach the boundary at $a=0$ at finite time, and more generally quantum wavefunctions have a finite probability to tunnel under the potential barrier to reach $a=0$.  Indeed, this phenomenon is associated with the Hartle-Hawking no-boundary proposal for the wavefunction of the universe \cite{Hartle:1983ai}, or perhaps more directly with the Vilenkin tunneling-from-nothing wavefunction \cite{Vilenkin:1982de,Vilenkin:1984wp}.

Of course, in the strict semiclassical limit $\ell_p\rightarrow 0$ such tunneling should not occur.  In this limit, we should be able to describe a general theory of quantum gravity by cylinder diagrams, which on Kaluza-Klein reduction to one-dimension are associated with some matter Hamiltonian $H_0$.  Since the above tunneling turns off in this limit, this limiting matter Hamiltonian should in fact be essentially self-adjoint.  In the example above, this might be because the semiclassical limit is associated with some preferred boundary condition that defines the essentially self-adjoint $H_0$ from $H$.  And it may be useful to explore the perturbative expansion in $\ell_p$, where a similar structure should arise at all orders.

However, at the non-perturbative level any notion of an essentially self-adjoint constraint defined on a single-universe Hilbert space may cease to be relevant.  It would be extremely interesting to understand the structure that replaces it, as well as the associated physics that results.

\paragraph{Higher-dimensional generalizations:}

A natural higher dimensional generalization of the ideas discussed above would be to consider two dimensional spacetimes, where the allowed boundaries are line segments or closed loops. As a prominent example, we may regard string theory as a two-dimensional theory of gravity, taking a perspective where we regard the worldsheet as spacetime. From this perspective, target space is merely a manifold for matter fields. Boundary conditions correspond to asymptotic states of open or closed strings. Many of the above constructions generalize naturally to such cases (see eg., \cite{Hawking:1991vs,Lyons:1991im} for related comments).

Such a two-dimensional `quantum gravity' could  arise even for rather prosaic systems by characterizing the dynamics in terms of surface-like degrees of freedom. For instance, Ising-like models with $\ZZ_2$-valued spins may described  terms of domain wall variables (the surfaces across which spins flip), cf., \cite{Polyakov:1987ez,Fradkin:1980gt,Iqbal:2020msy}. Likewise, the effective dynamics of QCD flux tubes is captured by a two-dimensional theory of the confining string \cite{Aharony:2013ipa,Dubovsky:2015zey}. These theories are gravitational in the sense that they sum over two-manifolds modulo diffeomorphisms (though the fields living on the manifolds  do not include a dynamical metric). In particular, the dynamics of the domain walls typically involves a sum over topologies. If interpreted as a two-dimensional theory of gravity living on the domain walls,  these theories clearly fit very naturally into the EST paradigm of \cref{sec:EST}.  We can then give boundary conditions by specifying loops in target space on which domain walls end. The amplitudes correspond in the statistical theory to correlation functions of `defect' loop operators ('t Hooft loops for the $\ZZ_2$ spin symmetry). This is analogous to the worldline description of a theory with particle-like excitations as a one-dimensional theory of gravity, where we integrate in the worldline metric. Integrating in two dimensional gravitational dynamics is significantly more involved. In the case of the effective QCD string, the Nambu-Goto dynamics may be recast as a $T\overline{T}$-deformed free boson theory, which could be interpreted as a two dimensional model coupled to gravity \cite{Dubovsky:2018bmo,Cardy:2018sdv,Callebaut:2019omt}.

Another example in this vein is given by the genus expansion of large-$N$ gauge theories, which one might hope to describe as a two-dimensional theory of gravity. This hope is realized concretely for $\mathcal{N}=4$ Yang-Mills via the AdS/CFT correspondence: we may describe this theory as two-dimensional gravity (a string theory), though the target space is $\target = \mathrm{AdS}_5\times \mathbf{S}^5$ rather than simply the four-dimensional spacetime on which the gauge theory resides. Since the target space contains dynamical gravity, gauge-invariant boundary conditions are associated with strings ending on the (fixed) asymptotic boundary of $\target$. For example, we have boundary conditions given by vertex operators corresponding to local operator insertions in the boundary CFT, and by loops on which strings end on the boundary corresponding to Wilson loops. This situation was described in \cite{McNamara:2020uza}. In particular, they interpreted JT gravity as a description of worldsheets in a topological string theory, and the corresponding amplitudes as correlation functions of Wilson loops in the dual Chern-Simons gauge theory. Again, this example fits naturally with the EST considerations of \cref{sec:EST}, and in particular as a higher-dimensional generalization of worldline descriptions of AdS/CFT noted at the end of that section. As presented there, the restriction of boundary conditions to the asymptotic boundary of AdS was rather artificial, but this restriction is expected to become a requirement of gauge invariance once our worldline particles include an interacting graviton.

Having noted the similarity between one-dimensional (worldline) and string theories, we should point out a notable difference. With the exception of the GAT theories in \cref{sec:GAT}, our constructions above were sensitive to the off-shell content of the matter theory: that is, we were not limited to states on the worldline satisfying the constraint $H=0$. This can be traced back to the boundary of the integral over the lapse $T$, corresponding to `short' worldlines with $T=0$. However, string theory is different in this respect: amplitudes are sensible only with on-shell string states  corresponding to physical vertex operators. This is ultimately due to the gauging of Weyl symmetry, via which any string state can be thought of as specified at infinite distance on the worldsheet. One can attempt to eschew stringy constructions and attempt to either quantize a particular class of diffeomorphism invariant two dimensional gravitational dynamics on two-surfaces, or quantize a gauge fixed system where we only consider rigid geometric structures. For example, we can take two-geometries to be cylinders and impose as in the GAT construction independent delta function constrains for time translations and spatial rotations (this bears some resemblance to ambitwistor string constructions \cite{Mason:2013sva,Adamo:2013tsa}). Whether such models make sense, and how one might interpret off-shell string field theory in the language described above, deserve further investigation.\footnote{We believe that Hilbert space of  conventional critical string theory is obtained from a QFT like construction for the worldsheet quantum gravity theory and thus does not have superselection sectors. On the other hand for non-critical strings it is unclear to us whether there exists a scheme that allows for a construction involving superselection sectors and baby universes. } A relevant perspective may be offered by the worldsheet description of out-of-equilibrium dynamics, as recently investigated in \cite{Horava:2020she,Horava:2020val}.

\paragraph{Spacetime D-branes:}
Above, we computed amplitudes defined by the path integral over all worldlines with fixed boundaries. A natural generalization allows the inclusion of \emph{dynamical} boundaries. This means that the worldline path integral includes a sum over configurations with additional boundaries of a specified type. We call such boundaries `spacetime D-branes', since they are analogous to D-branes in string theory considered from the perspective of the worldsheet. Similar boundaries have interesting and important effects in JT gravity and related models \cite{Saad:2019lba,Blommaert:2019wfy,Marolf:2020xie,Blommaert:2020seb}. How are such boundaries interpreted in the context of the worldline models discussed here?

Amplitudes with a spacetime D-brane can be described by including the exponential $\exp(\lambda \hat{b})$, where $\hat{b}$ is a boundary-inserting operator of the relevant type, and $\lambda$ a `coupling' specifying the amplitude for spacetime to end on the brane. From the perspective of the target-space path integral over fields $\Phi(x)$ (such as \eqref{eq:ESTQGA}), this becomes the insertion of an exponential $e^{\int J(x)\Phi(x)}$, which in QFT terms is a source for the field $\Phi$. In this way, a spacetime D-brane can be interpreted as a deformation of the field theory action, or correspondingly (at least in the Euclidean statistical models of section \ref{sec:EST}) as a deformation of the distribution over $\alpha$-states. Another natural example (if we  augment our worldine theory to describe gauge fields in target space) is a Wilson loop, which can be described as a spacetime D-brane whereby worldlines can end with fields taking values on a specified closed curve in target space.

Furthermore, one might consider the spacetime D-branes themselves to be dynamical objects, integrating over different possible choices of such boundaries. For instance, the Wilson lines in the example above may be interpreted as the trajectory of a heavy charged `probe' particle: we may promote this particle to be dynamical by an appropriate integration over spacetime D-branes corresponding to all possible heavy particle trajectories.  This gives one way to interpret the discussion of \cite{Hawking:1991vs}. A more careful consideration of such dynamical boundaries in worldline theories may help to elucidate their role more generally in higher-dimensional models.

%~~~~~~~~~~~~~~~~~~~~~~~~~~~~~~~~~~~~~~~~~~~~~~
\acknowledgments
%~~~~~~~~~~~~~~~~~~~~~~~~~~~~~~~~~~~~~~~~~~~~~~

We thank Tarek Anous, Steve Giddings, Seth Koren, Jorrit Kruthoff, and Raghu Mahajan for conversations motivating much of this work.  We also thank the participants of the 2020 KITP Gravitational Holography program for motivating questions during related talks.
EC and MR  were supported by  U.S. Department of Energy grant DE-SC0009999 and by funds from the University of California.
DM and HM were supported by NSF grant PHY1801805 and by funds from the University of California.  H.M.~was also supported in part by a DeBenedictis Postdoctoral Fellowship.

%%%%%%%%%%%%%%%%%%%%%%%%%%%%%%%%%%%%%%%%%%%%%%%%%%%%%%%%%%%%
\appendix

%~~~~~~~~~~~~~~~~~~~~~~~~~~~~~~~~~~~~~~~~~~~~~~~
\section{QFT positivity from Euclidean worldlines}
\label{app:WPD}
%~~~~~~~~~~~~~~~~~~~~~~~~~~~~~~~~~~~~~~~~~~~~~~

As noted in \cref{sec:EQFT}, positivity of the quantum field theory inner product is manifest from the path integral over fields, but it is less obvious from the worldline formalism. As noted in  \cite{Marolf:2020xie} one can attribute this to the fact that  it is unclear what conditions one ought to impose on the quantum gravity path integral for a positive-definite inner product. For a Gaussian field, one can however demonstrate the positive-definiteness of the full QFT Hilbert space directly from the worldline formalism. We outline an argument below, in case it is helpful in providing an insight into more general issues in quantum gravity.

Let us first rewrite the inner product in a slightly different way, using functions $F$ that solve the Klein-Gordon equation sourced by $f$:
\begin{equation}
	F(x) = \int_{\target} dy \,G(x,y)\, f(y) \iff (m^2-\nabla^2)F = f\,.
\end{equation}
We have (with the $\mathbb{Z}_2$ involution $\sigma$)
\begin{equation}\label{eq:IPF}
	\braket{f_2}{f_1} = \int_{\target} dx \, F_2^*(\sigma(x)) (m^2-\nabla^2) F_1(x).
\end{equation}
We now note that if $F_1$ is positive frequency, meaning that $f_1 =(m^2-\nabla^2)F_1 $ vanishes on $\target^-$, this inner product can be written as an integral over $\target^+$ only. But we can then integrate by parts, and if $F_2$ is positive frequency it means that
$F_2^*\circ\sigma$ is negative frequency; the resulting integrand then vanishes on $\target^+$. All that remains are boundary terms, and we can write the inner product as an integral on $\Sigma$ (the fixed point locus of $\sigma$).

To see this in more detail, let us introduce some notation. First, define a `symplectic form' $\Omega$ on the space of functions on ${\target}$ (or perhaps the space of functions that satisfy the Klein-Gordon equation $(m^2-\nabla^2)F=0$ in a neighborhood of $\Sigma$):
\begin{equation}
	\Omega(F_1,F_2) = i \int_\Sigma \left(F_2 \nabla_n F_1 - F_1\nabla_n F_2\right).
\end{equation}
Here $\nabla_n$ is the normal derivative at $\Sigma$, pointing out of $\target^+$ and into $\target^-$. The factor of $i$ is a nice convention because it disappears if we `Wick rotate' to Lorentzian signature, and this becomes the usual symplectic form on Cauchy data for the wave equation. From Stokes' theorem, we have
\begin{align}
	\Omega(F_1,F_2) &= i \int_{\target^+}(F_2\nabla^2 F_1 - F_1 \nabla^2 F_2) \\
	&= -i \int_{\target^+} F_2 (m^2-\nabla^2)F_1 \quad (F_2 \text{ negative frequency}).
\end{align}
In the last line we assume that $F_2$ is negative frequency, which means that $(m^2-\nabla^2)F_2=0$ on $\target^+$. This now resembles our inner product \eqref{eq:IPF}.

We can in fact write the precise relation in terms of the `complex structure' $J$ acting on our space of functions by
\begin{equation}
	J F(X)  = i \,F^*(\sigma(X))\,.
\end{equation}
This operator satisfies $J^2=-1$ and maps positive frequency functions to negative frequency functions and vice-versa. We now have
\begin{equation}
	\braket{f_2}{f_1} = \Omega(F_1,JF_2).
\end{equation}
Writing this out explicitly gives us our inner product in terms of data on $\Sigma$ as
\begin{equation}\label{eq:IPSigma}
	\braket{f_2}{f_1} = -\int_\Sigma \nabla_n (F_2^* F_1),
\end{equation}
where we have used the fact that $\Sigma$ is fixed by $\sigma$, and $\nabla_n (JF) = -J \nabla_n F$.

We can now use Stokes' theorem again to write this as an integral on $\target^-$:
\begin{align}
	\braket{f_2}{f_1} &= \int_{\target^-} \nabla^2 (F_2^* F_1) \\
	&= 2\int_{\target^-}\left[(\nabla F_2^*)\cdot (\nabla F_1)+ m^2 F_2^* F_1\right],
\end{align}
where we have used the fact that $F_1,F_2$ are positive frequency again to write $\nabla^2=m^2$. In this form, the inner product is manifestly positive semi-definite:
\begin{equation}
	\langle f|f\rangle = 2\int_{\target^-}\left(|\nabla F|^2+ m^2 |F|^2\right) \geq 0.
\end{equation}

Now that we have a Hilbert space, let's understand it a little better. First, we can think about it in terms of positive frequency functions $F$, satisfying $(m^2-\nabla^2)F=0$ on $\target^-$. This constraint determines $F$ from its values on $\target^+$ (existence and uniqueness for the Dirichlet problem given data on $\Sigma$), so the space of positive frequency $F$ is roughly the space of all functions on $\target^+$. But we have seen that the inner product can be written in terms of data on $\Sigma$ only, so the Hilbert space is much smaller: we have `null states' corresponding to functions $F$ which are nonzero only in the interior of $\Sigma$. From \eqref{eq:IPSigma} it looks like the relevant data is the value of $F$ and its normal derivative on $\Sigma$. But that's still too much data: $F$ and $\nabla_nF$ on $\Sigma$ are related (nonlocally) by the positive frequency condition. We should therefore think of the states in the single-universe Hilbert space as depending on a single (complex) function of $\Sigma$, which we might think of as the wavefunction of a single particle.

In terms of the source $f$, (a dense set of) the Hilbert space consists of functions with compact support $\operatorname{supp}(f)\subset \target^+$, modulo the image of $(m^2-\nabla^2)$ among such functions. The latter give null states, since functions $f=(m^2-\nabla^2)F$ with $F$ vanishing in a neighbourhood of $\Sigma$ have zero inner product with any state. Integrating by parts, we can see this as a statement of the Hamiltonian constraint or Wheeler-deWitt equation:
\begin{equation}
	(m^2-\nabla^2) \ket{\Phi(x) } =0.
\end{equation}
That is, $0= \ket{ (m^2-\nabla^2)\, F } =\int_{\target^+}  \left((m^2-\nabla^2)\, F\right) \Phi = \int_{\target^+} F (m^2-\nabla^2)\Phi $. This follows because $G(x,y)$ obeys the Laplacian away from coincident points, and the restriction on boundary conditions precludes contact terms.

For multiple insertions of $\Phi$, we have a Fock space built on this one-universe Hilbert space. However, this is not completely immediate, since the amplitudes contain contractions between insertions in $\target^+$, different from the one-universe inner product because there's no $\sigma$ reflection required. One way to proceed is by constructing `connected' states, inserting some $\Phi(x)$ but subtracting by hand all contractions within $\target^+$. Inner products between such connected states sum only over contractions between one `ket' insertion in $\target^+$ and one `bra' insertion in $\target^-$.

%%%%%%%%%%%%%%%%%%%%%%%%%%%%%%%%%%%%%%%%%%%%%%%%%%%%%%%%%%%%
% \bibliographystyle{JHEP}
% \bibliography{wlineqg-refs}

 \providecommand{\href}[2]{#2}\begingroup\raggedright\endgroup

\end{document}